\documentclass[fleqn,10pt]{SelfArx} % Document font size and equations flushed left

\definecolor{color1}{RGB}{0,0,90} % Color of the article title and sections
\definecolor{color2}{RGB}{0,20,20} % Color of the boxes behind the abstract and headings

\usepackage{hyperref} % Required for hyperlinks
\hypersetup{hidelinks,colorlinks,breaklinks=true,urlcolor=color2,citecolor=color1,linkcolor=color1,bookmarksopen=false,pdftitle={Title},pdfauthor={Author},urlcolor=blue}

\usepackage{graphicx}
\usepackage[square]{natbib}
\usepackage{amsmath}
\usepackage{multirow}
\usepackage{subfigure}

\newcommand{\n}[1]{\mathrm{#1}}

\JournalInfo{Published in International Journal of Refrigeration, Vol. 34 (8), 1805-1816, 2011} % Journal information
\Archive{\href{http://dx.doi.org/10.1016/j.ijrefrig.2011.05.021}{DOI: 10.1016/j.ijrefrig.2011.05.021}} % Additional notes (e.g. copyright, DOI, review/research article)

\PaperTitle{Determining the minimum mass and cost of a magnetic refrigerator} % Article title

\Authors{R. Bj\o{}rk, A. Smith, C. R. H. Bahl and N. Pryds} % Authors
\affiliation{\textit{Department of Energy Conversion and Storage, Technical University of Denmark - DTU, Frederiksborgvej 399, DK-4000 Roskilde, Denmark}} % Author affiliation
\affiliation{*\textbf{Corresponding author}: rabj@dtu.dk} % Corresponding author

 % Keywords - if you don't want any simply remove all the text between the curly brackets
\newcommand{\keywordname}{Keywords} % Defines the keywords heading name

\Abstract{An expression is determined for the mass of the magnet and magnetocaloric material needed for a magnetic refrigerator and these are determined using numerical modeling for both parallel plate and packed sphere bed regenerators as function of temperature span and cooling power. As magnetocaloric material Gd or a model material with a constant adiabatic temperature change, representing a infinitely linearly graded refrigeration device, is used. For the magnet a maximum figure of merit magnet or a Halbach cylinder is used. For a cost of \$40 and \$20 per kg for the magnet and magnetocaloric material, respectively, the cheapest 100 W parallel plate refrigerator with a temperature span of 20 K using Gd and a Halbach magnet has 0.8 kg of magnet, 0.3 kg of Gd and a cost of \$35. Using the constant material reduces this cost to \$25. A packed sphere bed refrigerator with the constant material costs \$7. It is also shown that increasing the operation frequency reduces the cost. Finally, the lowest cost is also found as a function of the cost of the magnet and magnetocaloric material.}

\begin{document}

\flushbottom % Makes all text pages the same height

\maketitle % Print the title and abstract box

%\tableofcontents % Print the contents section

\thispagestyle{empty} % Removes page numbering from the first page

%\section*{Nomenclature}
%\begin{table}[!hb]
%\begin{tabular}{|ll|}
%  \hline
%  \textit{Variables} &                                                \\
%  $\Lambda_\mathrm{cool}$   & Magnet efficiency parameter for magnetic refrigeration [T$^{2/3}$] \\
%  $M^{*}$                   & Magnet efficiency parameter (-) \\
%  $H$                       & Magnetic field (A m$^{-1}$) \\
%  $V$                       & Volume (m$^3$) \\
%  $B_\mathrm{rem}$          & Remanence (T) \\
%  $\dot{Q}$                 & Cooling power (W) \\
%  $T_\mathrm{span}$         & Temperature span (K) \\
%  $\dot{Q}_\mathrm{max}$    & Maximum cooling power (W) \\
%  $T_\mathrm{span, max}$    & Maximum (no load) temperature span (K) \\
%  $m$                       & Mass (kg) \\
%  $\Delta{}x$               & Fluid stroke length (\%) \\
%  $T$                       & Temperature (K) \\
%  $h$                       & Height (mm) \\
%  $d_\n{p}$                 & Particle diameter (mm) \\
%  $f$                       & Frequency (Hz) \\
%  \hline
%  \end{tabular}
%\end{table}
%
%\begin{table}[!hb]
%\begin{tabular}{|ll|}
%  \hline
%  \textit{Greek} &                                              \\
%  $\mu_0$                   & Permeability of free space (m kg s$^{-2}$ A$^{-2}$) \\
%  $\rho$                    & Density (kg m$^{-3}$) \\
%  $\epsilon$                & Porosity (-) \\
%  $\tau_\n{rel}$            & Relative cycle time (-) \\
%  \hline
%\end{tabular}
%\end{table}
%\clearpage

\section{Introduction}
Magnetic refrigeration is a new environmentally friendly cooling technology with a potential for high energy efficiency. The technology is based on the magnetocaloric effect (MCE), which is the temperature change that most magnetic materials exhibit when subjected to a changing magnetic field. For the benchmark magnetocaloric material (MCM) used in magnetic refrigeration, gadolinium, the adiabatic temperature change is no more than 4 K in a magnetic field of 1 T \citep{Dankov_1998,Bjoerk_2010d}, and therefore a magnetic refrigeration device has to utilize a regenerative process to produce a large enough temperature span to be useful for refrigeration purposes. The most utilized process for this is called active magnetic regeneration (AMR) \citep{Barclay_1982}.

An AMR consists of a porous matrix of a solid magnetocaloric material and a heat transfer fluid that can flow through the matrix and reject or absorb heat. The solid matrix is termed the regenerator. The heat is then transferred to a cold and hot heat exchanger at either end of the AMR. Using this system a temperature gradient can be built up that can be much larger than the adiabatic temperature change produced by the magnetocaloric material. Typically the porous matrix is either a packed sphere bed \citep{Okamura_2005,Tura_2009} or consists of parallel plates \citep{Zimm_2007, Bahl_2008}. A review of different magnetic refrigeration devices is given in \citet{Yu_2010}.

The temperature span and cooling power generated by an AMR device depends on the process parameters specific to each AMR system. These are the shape and packing of the magnetocaloric material, the temperature of the surroundings and the properties of the MCM used, as well as the properties of the heat transfer fluid, flow system, magnetic field, geometry of the AMR etc. In operation, two performance parameters are of key importance, the temperature span, $T_\n{span}$, which is the difference between the temperature of the hot and the cold reservoir at either end of the AMR, $T_\mathrm{hot}$ and $T_\mathrm{cold}$, respectively, and the cooling power, $\dot{Q}$, generated by the AMR. For given process parameters, $T_\n{span}$ and $\dot{Q}$ trace out a curve called the cooling curve. In a ($T_\n{span}$,$\dot{Q}$) diagram the cooling curve is in many cases of interest approximated by a straight line going from $(0,\dot{Q}_\n{max})$ to $(T_\n{span,max},0)$ with a negative slope. As the maximum cooling power and the maximum temperature span cannot be realized at the same time, the operation point will lie on the cooling curve somewhere in between the two extrema. Neither extremity of the curve is of interest for actual operation.

Determining the cost of an AMR is of general interest in order to evaluate the cost-performance of the technology. An assessment of the costs for a residential air conditioner based on magnetic cooling presented in \citet{Russek_2006} concluded that the cost of the magnet is of great importance and furthermore found that the cost of the magnet and magnetocaloric material for such a magnetic air conditioner can be competitive with conventional air conditioners. It has also been investigated \citep{Egolf_2007} whether magnetic heat pumps can compete with conventional heat pumps. Based on simple theoretical calculations it is estimated that magnetic heat pumps are only 30\% more expensive than conventional heat pumps. However, the price of the magnet is never considered in this analysis. The total cost of a AMR magnetic refrigeration device has recently been considered in \citet{Rowe_2009} and \citet{Rowe_2010a} who defined a general performance metric for active magnetic regenerators. The cost and effectiveness of the magnet design is included in this metric as a linear function of the volume of the magnet, and the generated field and the amount of magnetocaloric material used is also included in the metric. However, the metric has to be calculated for a specific refrigeration system and can not be used to predict the general performance per dollar of the magnetic refrigeration technology. A figure of merit used to evaluate the efficiency of a magnet design used in magnetic refrigeration has been introduced in \citet{Bjoerk_2008} but this does not take the performance of the actual AMR system into account.

Here, we are interested in determining the lowest combined cost of magnet and magnetocaloric material needed for a magnetic refrigerator as a function of a desired temperature span and cooling power. Determining the lowest combined cost of magnet and magnetocaloric material allows for the determination of the major source of cost of a magnetic refrigeration device and thus allows the technology to be compared to competing refrigeration technologies. Note that we wish to determine the lowest combined cost, i.e. the cost of the materials for the cheapest magnetic refrigerator. Thus whenever a cost is given in this article it is the cost of the materials needed to construct the device that is meant. It is important to state that the cheapest system might not be the most efficient device possible, i.e. the device that has the highest coefficient of performance (COP). However, such a device would use more magnet and magnetocaloric material than the lowest cost device. The overall lifetime cost of a magnetic refrigeration device include both capital cost and operating cost but this is not considered in the present analysis. %and thus the mass and cost calculated here are the minimum values for a magnetic refrigeration device. Thus the analysis presented here discuss the minimum cost of the materials needed for the construction of a magnetic refrigeration and ignores the operating cost.

\section{Determining the mass of the magnet}
A magnetic refrigerator in which the magnetic field is provided by a permanent magnet assembly, as is the case for almost all magnetic refrigeration devices \citep{Bjoerk_2010b}, is considered. A measure of the efficiency of a magnet used in magnetic refrigeration is given by the $\Lambda_\n{cool}$ parameter, as defined in \citet{Bjoerk_2008}. The  $\Lambda_\n{cool}$ parameter is defined as
\begin{eqnarray}
\Lambda_\mathrm{cool} \equiv \left(\langle B^{2/3}\rangle - \langle B^{2/3}_{\mathrm{out}}\rangle \right)\frac{V_{\mathrm{field}}}{V_{\mathrm{mag}}}P_{\mathrm{field}}~,
\end{eqnarray}
where $V_{\mathrm{mag}}$ is the volume of the magnet(s), $V_{\mathrm{field}}$ is the volume where a high flux density is generated, $P_{\mathrm{field}}$ is the fraction of an AMR cycle that magnetocaloric material is placed in the high flux density volume, $\langle B^{2/3}\rangle$ is the volume average of the flux density in the high flux density volume to the power of 2/3 and $\langle B^{2/3}_{\mathrm{out}}\rangle$ is the volume average of the flux density to the power of 2/3 in the volume where the magnetocaloric material is placed when it is being demagnetized. Note that it is the magnetic flux density generated in an empty volume that is considered, i.e. $\mathbf{B}=\mu_0\mathbf{H}$, and thus it is equivalent to speak of the magnetic flux density or the magnetic field.

A high $\Lambda_\mathrm{cool}$ generally favors small magnetic fields. However, this decreases the rate of heat transfer between the magnetocaloric material and the heat transfer fluid, ultimately decreasing the performance of the device. The optimum magnet configuration will reflect a trade-off between high magnet efficiency and high rates of heat transfer. This, in turn, will depend on the detailed configuration of the device. A number of magnet arrays for magnetic refrigeration have been compared in \citet{Bjoerk_2010b}.

An alternative way to classify permanent magnet arrays is to consider the so-called figure of merit, $M^*$, which is defined as \citep{Jensen_1996}
\begin{eqnarray}\label{Eq.Mstar_definition}
M^{*}=\frac{\int_{V_\n{field}}||\mu_{0}\mathbf{H}||^2dV}{\int_{V_\n{mag}}||\mathbf{B_\n{rem}}||^2dV}
\end{eqnarray}
where $V_\n{field}$ is the volume of the air gap where the desired magnetic field, $\mu_{0}\mathbf{H}$, is created and $V_\n{mag}$ is the volume of the magnets which have a remanence $\mathbf{B_\n{rem}}$. For isotropic materials with linear demagnetization characteristics this is a measure of the magnetic field energy in the air gap, divided by the maximum amount of magnetic energy available in the magnet material. Although magnet arrays with a high value of $M^*$ are not necessarily good magnets for magnetic refrigeration, $M^*$ offers a succinct way of characterizing the field strength attained in the high field region. It has the added advantage that an upper bound is known: It can be shown that the maximum value of $M^*$ is $0.25$. Here such a magnet will be termed the $M_{25}$ magnet. For the remainder of this paper we will characterize the magnet array using the figure of merit.

For specific permanent assemblies it is possible to calculate $M^*$ analytically; this will be considered later. Furthermore, if we limit ourselves to three-dimensional structures with a constant magnitude of the remanence (whose direction is allowed to vary) which generate a constant magnetic field in the gap, Eq. (\ref{Eq.Mstar_definition}) can be rewritten as
\begin{eqnarray}
M^{*}=\left(\frac{\mu_{0}H}{B_\n{rem}}\right)^2\frac{V_\n{field}}{V_\n{mag}}~.
\end{eqnarray}

Rearranging this equation by substituting the volume of the high field region for the mass of magnetocaloric material, $m_\n{mcm, field}$, divided by the mass density of the MCM, $\rho_\n{mcm}$, times one minus the porosity of the regenerator (including support structure), $(1-\epsilon{})$, and substituting the volume of the magnet by the mass, $m_\n{mag}$, divided by the density, $\rho_\n{mag}$, yields
\begin{eqnarray}\label{Eq.V_mag_with_m_mcm_pre}
m_\n{mag}=\left(\frac{\mu_{0}H}{B_\n{rem}}\right)^2\frac{m_\n{mcm, field}\rho{}_\n{mag}}{(1-\epsilon{})\rho{}_\n{mcm}M^{*}}~.
\end{eqnarray}
All terms that are a function of the magnetic field are on one side of the equation and thus from this equation we can calculate the mass of the magnet needed for a magnetic refrigerator, if we know the mass of magnetocaloric material as a function of $\mu_{0}H$ required to provide the desired temperature span, $T_\n{span}$, and cooling power, $\dot{Q}$. Also, $M^{*}$ as a function of $\mu_{0}H$ must of course also be known.

Note that the masses that are related in Eq. (\ref{Eq.V_mag_with_m_mcm_pre}) are the mass of the magnet and the mass of magnetocaloric material that is placed inside the magnet during an AMR cycle. If an symmetric AMR cycle and only a single regenerator is used the magnet will only be in use half of the cycle time, which is very inefficient. However, if one uses two regenerators, run completely out of phase and using the same magnet, double the amount of cooling power will be produced, of course using double the amount of magnetocaloric material but using the same amount of magnet. This system, in which the magnet is utilized at all times, is the most efficient system possible, and thus allows for a modification of Eq. (\ref{Eq.V_mag_with_m_mcm_pre}) such that
\begin{eqnarray}\label{Eq.V_mag_with_m_mcm}
m_\n{mag}=\frac{1}{2}\left(\frac{\mu_{0}H}{B_\n{rem}}\right)^2\frac{m_\n{mcm}\rho{}_\n{mag}}{(1-\epsilon{})\rho{}_\n{mcm}M^{*}}~.
\end{eqnarray}
where $m_\n{mcm}$ is now the total amount of magnetocaloric material, and it is assumed that the magnet is in use, i.e. filled with magnetocaloric material with a mass of $m_\n{mcm}/2$, at all times. AMR devices in which the magnet is in use almost continuously have previously been demonstrated \citep{Tusek_2010,Bjoerk_2010c}. Such an AMR which uses the magnet at all times is what is considered in the following.

As an example assume that the densities of the magnetocaloric material and the magnet are identical and that the system has a porosity of 0.5. Also, consider a magnetic field with the value of the remanence. For the magnet with the largest possible figure of merit, $M^* = 0.25$, i.e. the $M_{25}$ magnet, we obtain from Eq. (\ref{Eq.V_mag_with_m_mcm}) that the mass of the magnet must be four times the mass of magnetocaloric material used if the magnet is used at all times. If only a single regenerator was used then the mass of the magnet would be eight times that of the magnetocaloric material. However, for specific regenerator geometries and magnetocaloric material the mass of the magnet can be calculated more precisely.

\section{Determining the minimum mass of magnetocaloric material}
In order to use Eq. (\ref{Eq.V_mag_with_m_mcm}) to calculate the cost of a magnetic refrigeration system we need to know the mass of magnetocaloric material as function of $\mu_{0}H$ required to provide the desired temperature span, $T_\n{span}$, and cooling power, $\dot{Q}$. In order to determine this a parameter survey has been conducted where the cooling power has been computed using a numerical model for two different magnetic refrigeration devices, both using Gd modeled using the mean field theory (MFT) \citep{Morrish_1965}. The model used is a publicly available one-dimensional numerical model \citep{Engelbrecht_2006}. By varying the Nusselt-Reynolds correlations the model is capable of modeling both packed bed and parallel plate regenerators. For both regenerator geometries the model has previously been compared with both experimental data and other numerical models \citep{Engelbrecht_2008,Petersen_2008b,Bahl_2008}. In the numerical model, the temperature span is an input parameter and the cooling power is calculated for the specified process parameters. The governing equations of the model for the packed sphere bed and the parallel plate cases are given in \citet{Engelbrecht_2006} and \citet{Petersen_2008b}. The Gd material has a Curie temperature of $T_\n{c} = 293.6$ K and properties as given in \citet{Petersen_2008a}.

As previously mentioned the performance of the AMR depends on a number of process parameters, which are different for a parallel plate and a packed sphere bed regenerator. Here we consider a variety of process parameters, all of which have been chosen to span realistic values. A number of common process parameters which are shared between the parallel plate and the packed bed models have been fixed during the numerical experiments considered here. These are the length of the modeled regenerator which is taken to be 50 mm and the heat transfer fluid which is taken to be water with constant properties as given in \citet{Petersen_2008b}. Also, the temperature of the hot end of the AMR is kept fixed at $T_\n{hot}=298$ K and only a symmetric AMR cycle is considered. Other common process parameters include the cycle frequency, $f$, the relative cycle time, $\tau_\n{rel}$, which is the ratio between the time used for magnetization or demagnetization of the AMR and the time used for fluid displacement, the fluid stroke length, $\Delta{}x$, which describes the fraction of fluid that is displaced, the temperature of the cold end, $T_\n{cold}$ and the maximum magnetic field, $\mu_{0}H$. However, these parameters have been varied independently for the parallel plates and the packed bed cases. For both regenerator types a magnetic field with a temporal width of 55\% of a flow cycle and a temporal width of maximum field time of 45\% of the cycle time is used \citep{Bjoerk_2011a}. Thus the time it takes to ramp the magnetic field from 0 to the maximum value is 5\% of the cycle time. The magnetic field is ramped up at the start of the AMR cycle.

For the packed sphere bed regenerator the process parameters are the particle size, $d_\n{p}$, and the porosity, $\epsilon$. For a randomly packed sphere bed regenerator used in magnetic refrigeration the latter for a number of recently published systems is near 0.36 \citep{Okamura_2005,Jacobs_2009, Tura_2009} and therefore this parameter is fixed. The different process parameters are listed in Table \ref{Table.Packed_bed}. The values for the particle size, $d_\n{p}$, have been chosen based on reported experimental values \citep{Okamura_2005,Engelbrecht_2007a,Tura_2009}. The total number of parameter sets considered is 15876.

\begin{table}[t]
\begin{center}
\caption{The packed sphere parameters varied.}\label{Table.Packed_bed}
\begin {tabular}{lll}
Parameter & Values & Unit\\ \hline
$\Delta{}x$       & 70, 90, 110, 135, 150, 180, 215 & [\%]\\ %71 89 107 133 151 178 213
%$\tau_\n{tot}$    & 0.1, 0.25, 0.5, 1 & [s]\\
$f$               & 1, 2, 4, 10 & [Hz]\\
$\tau_\n{rel}$    & 0.1, 0.25, 0.5 & [-]\\
$d_\n{p}$         & 0.1, 0.25, 0.5 & [mm]\\
$\mu_{0}H$        & 0.4, 0.6, 0.8, 1.0, 1.2, 1.4, 1.6 & [T]\\
$T_\n{cold}$      & 268, 272, 276, 278, 280, 284 & \\
                  & 288, 292, 296, 298 & [K]
\end {tabular}
\end{center}
\end{table}

For the parallel plate regenerator two process parameters must be specified. These are the height of the fluid channel, $h_\mathrm{fluid}$, and the height of the plate, $h_\mathrm{plate}$. These have been chosen based on realistic experimental values \citep{Bahl_2008,Engelbrecht_2010}. The different process parameters considered for the parallel plate case are listed in Table \ref{Table.Parallel_plate}. However, for the parallel plate regenerator model a comparison with a two-dimensional AMR model leads to the requirement that a ``1D correctness'' parameter, $\Gamma$, must be much greater than one for the one-dimensional model to produce comparable results to a two-dimensional model \citep{Petersen_2008b}. Here all process parameters with $\Gamma < 3$ will not be considered further. These are process parameters with large values of $h_\n{fluid}$ and large values of $f$. Therefore the total number of parameter sets considered is 14994.

\begin{table}[t]
\begin{center}
\caption{The parallel plate parameters varied.}\label{Table.Parallel_plate}
\begin {tabular}{lll}
Parameter & Values & Unit\\ \hline
$\Delta{}x$      & 40, 50, 60, 70, 80, 90 & [\%]\\
%$\tau_\n{tot}$   & 0.25, 0.5, 1.0, 3.0, 6.0 & [s]\\
$f$              & 0.167, 0.33, 1, 2, 4 & [Hz]\\
$\tau_\n{rel}$   & 0.25, 0.50 & [-]\\
$h_\n{fluid}$    & 0.1, 0.25, 0.5 & [mm]\\
$h_\n{plate}$    & 0.1, 0.25, 0.5 & [mm]\\
$\mu_{0}H$       & 0.4, 0.6, 0.8, 1.0, 1.2, 1.4, 1.6 & [T]\\
$T_\n{cold}$     & 268, 273, 278, 283, 288, 293, 298 & [K]
\end {tabular}
\end{center}
\end{table}

For each of the set of process parameters the cooling power is calculated. Using this data, the mass of magnetocaloric material can be directly determined as a function of $\mu_{0}H$ for a desired $\dot{Q}$ and $T_\n{span}$ by only assuming that the cooling power is directly proportional to the mass of magnetocaloric material. For each temperature span and magnetic field the combined cost of the magnet and the MCM is simply calculated using Eq. (\ref{Eq.V_mag_with_m_mcm}) for all process parameters and the lowest cost selected. In the following we take the cost of the magnet material to be \$40 per kg and the cost for the magnetocaloric material to be \$20 per kg, similarly to \citet{Rowe_2010a}. The cost of assembly of the magnet and the regenerator is not included; although these costs may be substantial for the initial market entry devices, for a mass-produced product they are expected to be relatively minor. Differing cost estimates for the materials will be discussed subsequently. Using these numbers the total cost is calculated by adding the cost of the magnet and magnetocaloric material and minimizing this value for all process parameters. Note that, as argued previously, substantially more magnet compared to magnetocaloric material must be used. Therefore the calculation of the total cost is not very sensitive to the cost of the magnetocaloric material, but will scale roughly linearly with the cost of the magnet material. %Note also that, as we are interested in determining the lowest combined cost of magnet and magnetocaloric material the set of process parameter chosen might not result in the device which has the highest coefficient of performance (COP). However, such devices must use more magnet and magnetocaloric material than the lowest cost device and thus the mass and cost calculated here are the minimum values for a magnetic refrigeration device.

\section{Cost of a Gd AMR}
In order to determine the lowest combined cost of magnet and magnetocaloric material needed to produce a given desired temperature span and cooling power certain parameters must be specified. Here, we consider magnets with a remanence of 1.2 T, which is a common value for NdFeB magnets, which are the most powerful magnets commercially available today. These have a density of $\rho_\n{mag} = 7400$ kg/m$^3$. The density of Gd is $\rho_\n{mcm}=7900$ kg/m$^3$. For the parallel plate regenerators the porosity of the regenerator is calculated as $\epsilon = h_\n{fluid}/(h_\n{fluid}+h_\n{plate})$, while for the packed sphere bed the porosity is constant at $\epsilon = 0.36$. Note that none of these values for the porosity includes any support or housing structure for the regenerator.

\begin{figure}
\begin{center}
\includegraphics[width=1.0\columnwidth]{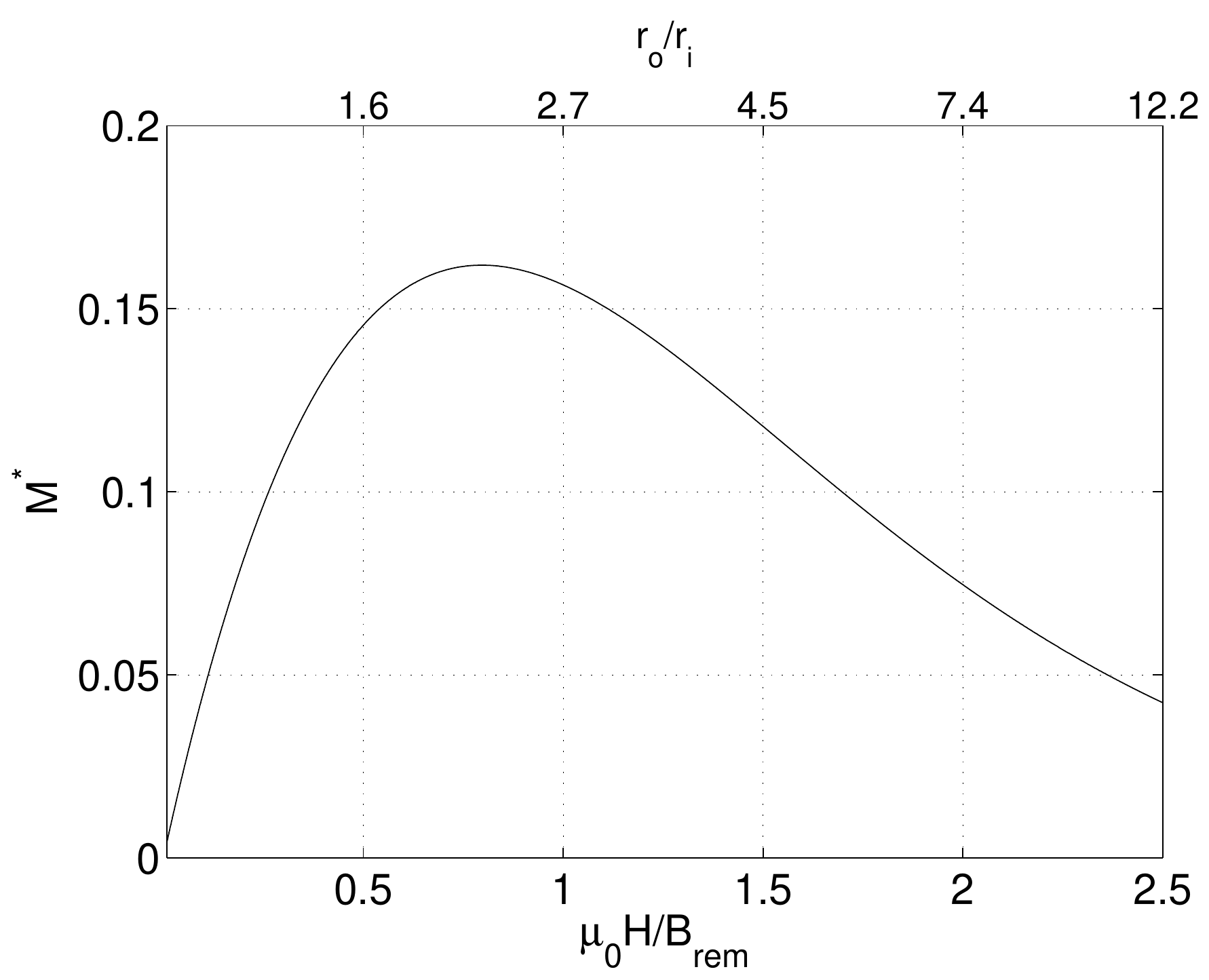}
\end{center}
\caption{The figure of merit, $M^{*}$, as a function of magnetic field in units of the remanence for a Halbach cylinder of infinite length.}
  \label{Fig.Mstar_Halbach}
\end{figure}

As previously mentioned for an $M_{25}$ magnet the figure of merit is $M^{*}=0.25$ for all values of $\mu_{0}H$. However, we will also consider the Halbach cylinder \citep{Mallinson_1973,Halbach_1980} which is a magnet design that has previously been used extensively in magnetic refrigeration devices \citep{Lu_2005,Tura_2007,Engelbrecht_2009,Kim_2009}. For this magnet design the efficiency parameter can be found analytically for a cylinder of infinite length, through the relation for the field in the cylinder bore, $\mu_{0}H = B_\n{rem}\n{ln}\left(\frac{r_\n{o}}{r_\n{i}}\right)$, where $r_\n{i}$ and $r_\n{o}$ are the inner and outer radius of the Halbach cylinder, respectively. Using this relation one gets \citep{Coey_2003}
\begin{eqnarray}\label{Eq.Mstar_Halbach}
M^{*} = \frac{\mathrm{ln}\left(\frac{r_\n{o}}{r_\n{i}}\right)^2}{\left(\frac{r_\n{o}}{r_\n{i}}\right)^{2}-1} = \frac{\left(\frac{\mu_{0}H}{B_\n{rem}}\right)^2}{e^{2\frac{\mu_{0}H}{B_\n{rem}}}-1}~.
\end{eqnarray}
This function is shown in Fig. \ref{Fig.Mstar_Halbach} and has an optimal value of $M^* \approx 0.162$ for a value of $\mu_{0}H/B_\n{rem}\approx 0.80$. For a Halbach of finite length the efficiency is lowered, depending on the length and inner radius of the device \citep{Bjoerk_2011b}. Here, for simplicity, we will only consider a Halbach cylinder of infinite length.

\subsection{A parallel plate regenerator of Gd}
\begin{figure}[!t]
\begin{center}
\subfigure[a]{\includegraphics[width=0.44\textwidth]{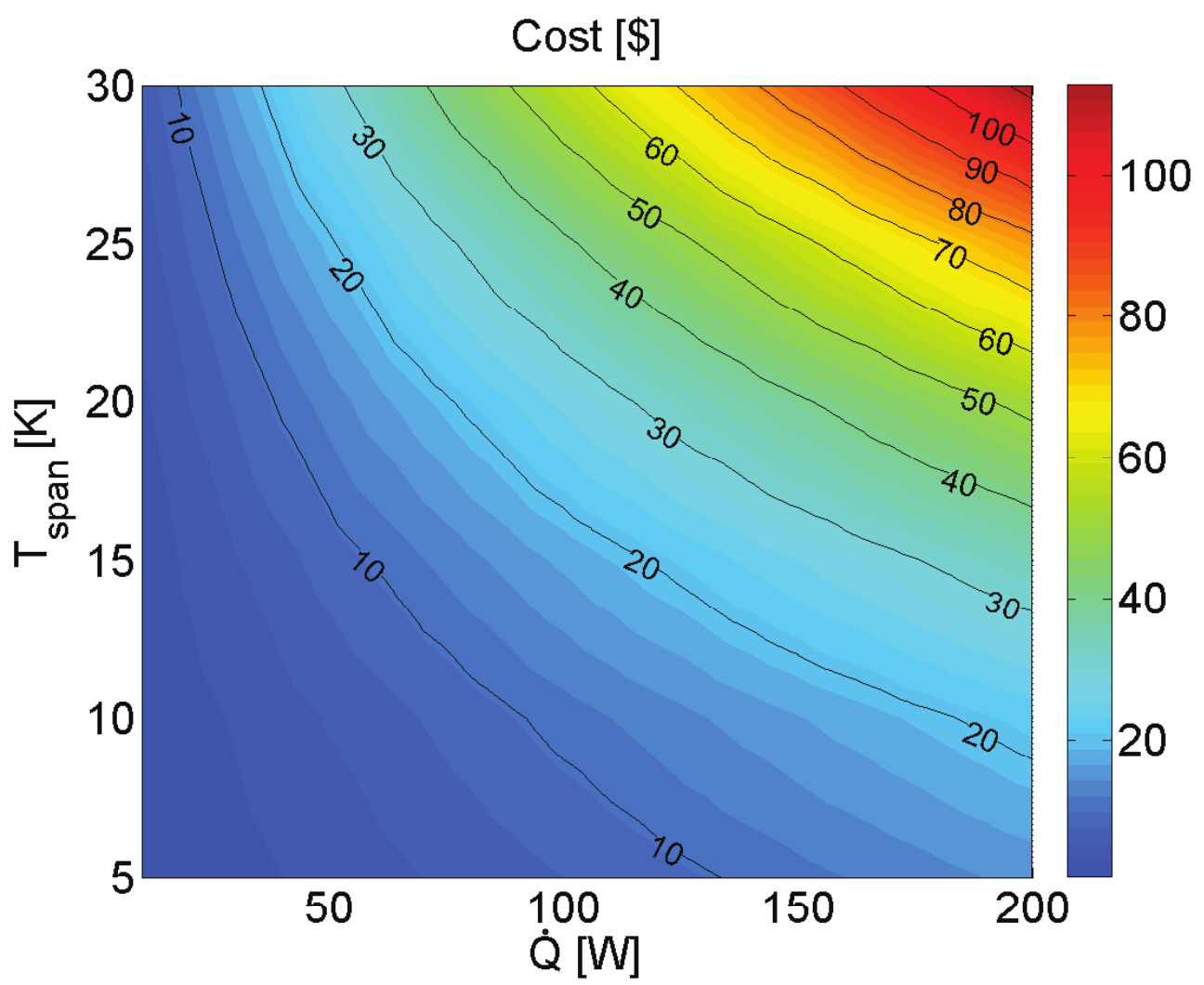}}
\subfigure[b]{\includegraphics[width=0.44\textwidth]{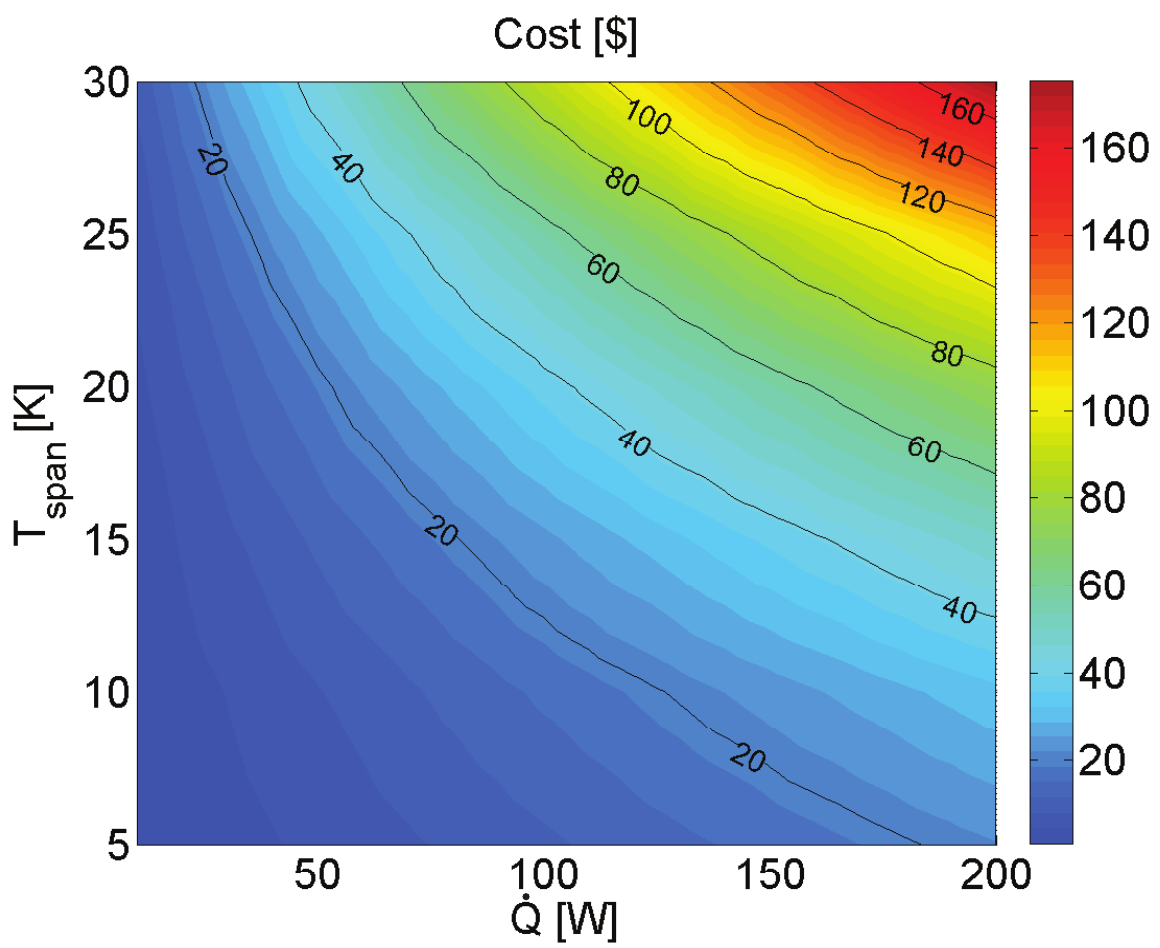}}
\end{center}
    \caption{The minimum cost in \$ of a parallel plate magnetic refrigeration system of Gd as a function of temperature span and cooling power for (a) a $M_{25}$ permanent magnet assembly and (b) a Halbach cylinder of infinite length.}
    \label{Fig.cost}
\end{figure}

\begin{figure}
\begin{center}
\subfigure[a]{\includegraphics[height=0.73\columnwidth]{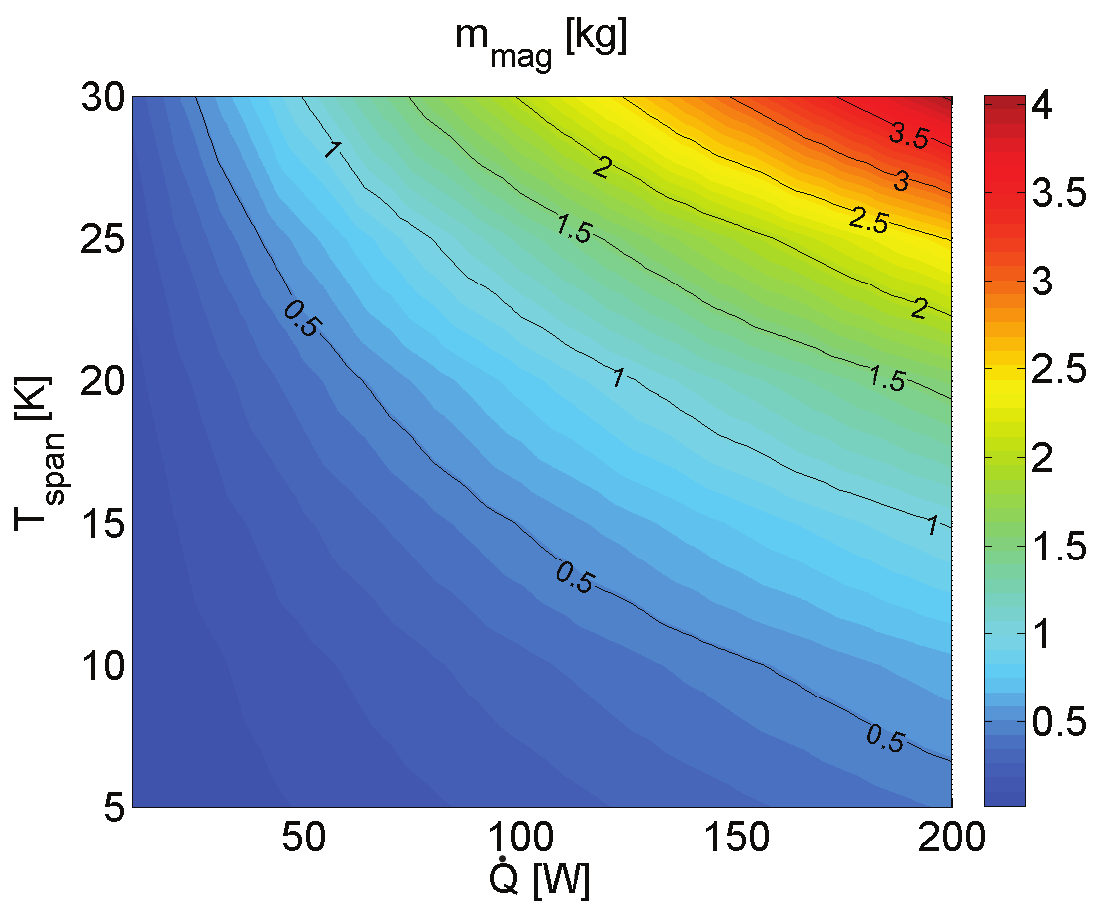}}
\subfigure[b]{\includegraphics[height=0.73\columnwidth]{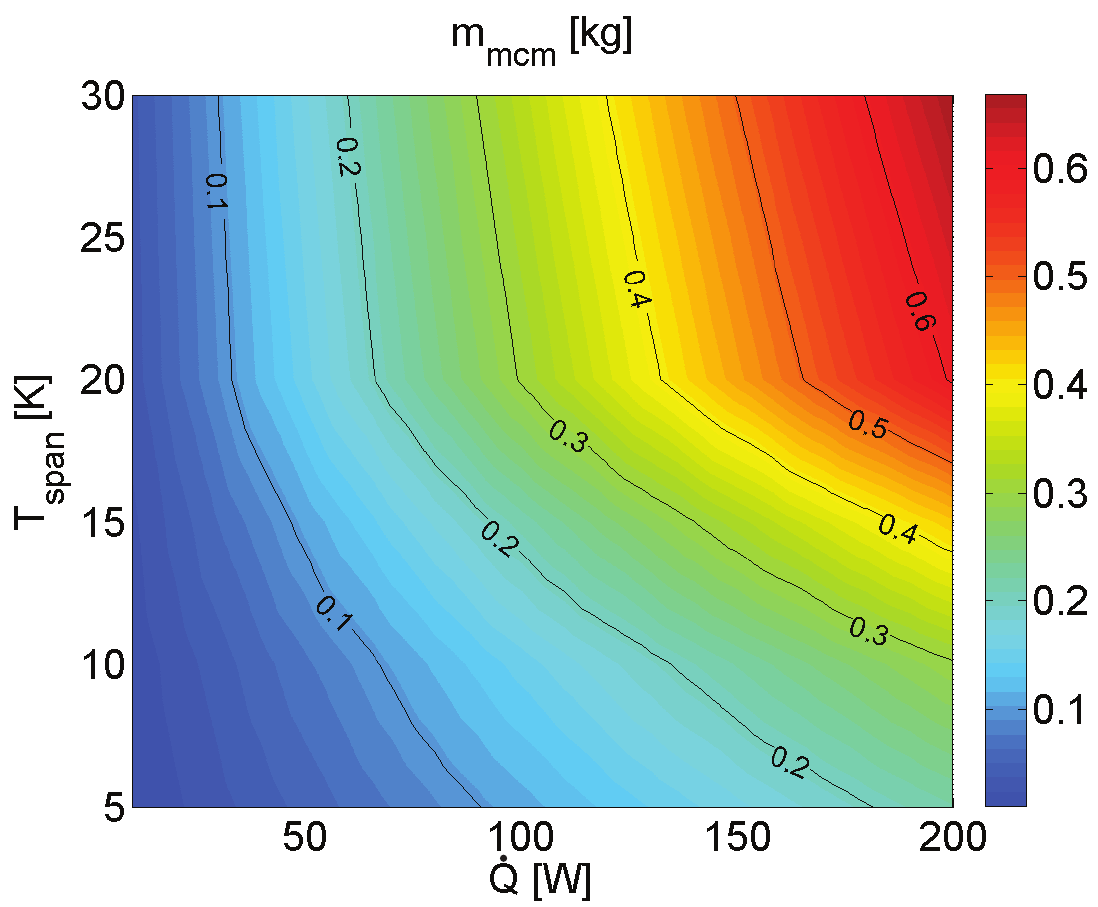}}
\subfigure[c]{\includegraphics[height=0.73\columnwidth]{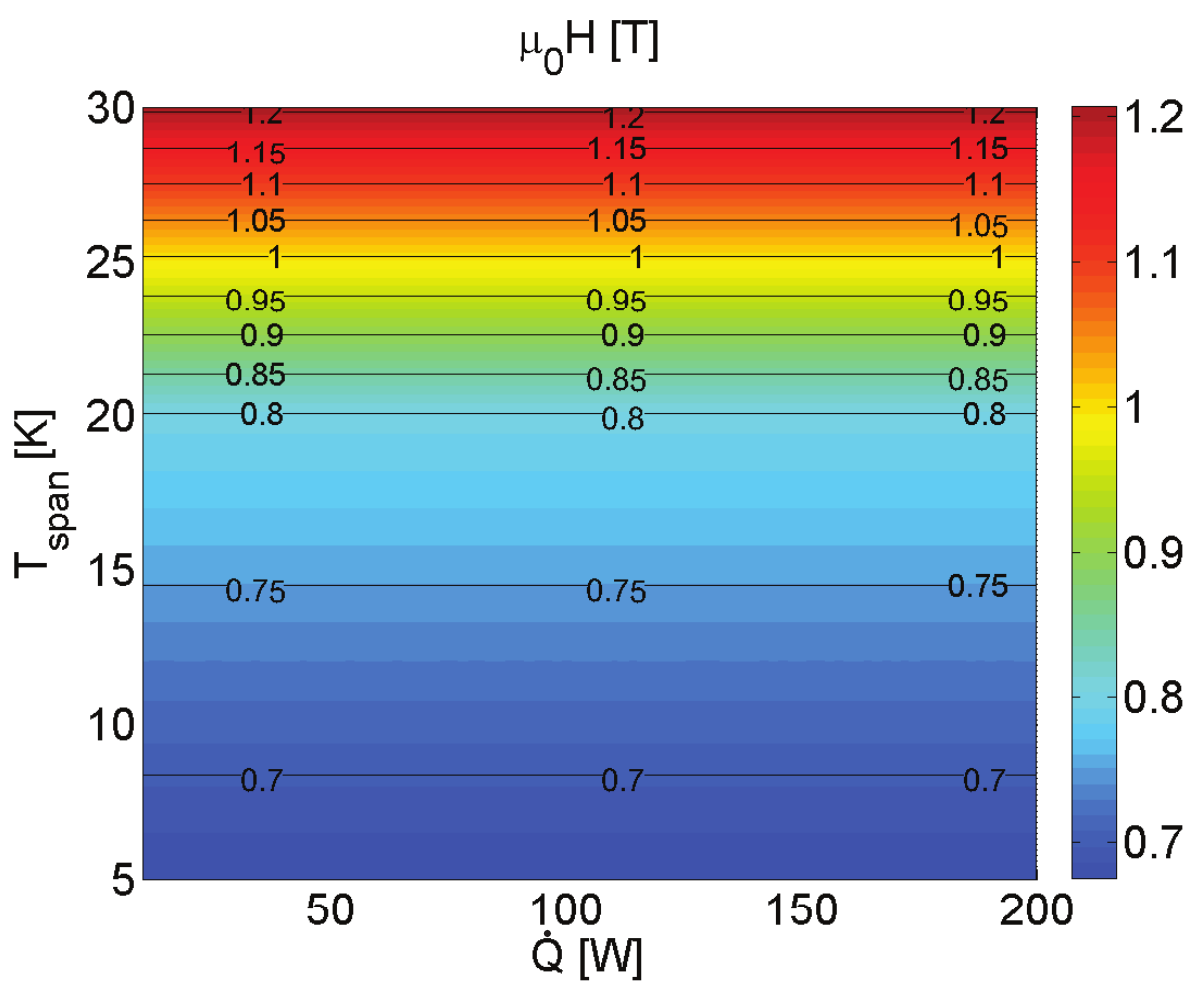}}
\end{center}
\caption{(a) The amount of magnet material, (b) the corresponding amount of magnetocaloric material and (c) the corresponding magnetic field for the parallel plate magnetic refrigeration system using Gd with the lowest combined cost of the magnet and magnetocaloric material for a Halbach cylinder of infinite length, i.e. with $M^*$ given by Eq. (\ref{Eq.Mstar_Halbach}).}
    \label{Fig.Halbach_com}
\end{figure}

We begin by analyzing the cost of a parallel plate regenerator of Gd, as this is the benchmark system in magnetic refrigeration.  Using the approach described above the total minimum combined cost of a parallel plate Gd regenerator with an $M_{25}$ magnet as a function of desired temperature span and cooling power of the AMR is shown in Fig. \ref{Fig.cost}a while for a Halbach magnet the minimum cost is shown in Fig. \ref{Fig.cost}b for the process parameters considered here. The corresponding amount of magnet material, magnetocaloric material and magnetic field for the minimum cost device using a Halbach magnet are shown in Fig. \ref{Fig.Halbach_com}. Here we are only interested in determining the minimum cost of a magnetic refrigeration device, and thus the process parameters for the lowest cost device will not be analyzed.

From Fig. \ref{Fig.cost} we see that the cost of a refrigeration system increases with both temperature span and cooling power and that a device using a Halbach cylinder is $\sim{}25-50\%$ more expensive than when an $M_{25}$ magnet is used. For example a system that produces 100 W of continuous cooling at a temperature span of 20 K using a Halbach magnet will have a minimum cost of \$35.

In Fig. \ref{Fig.Halbach_com} we see that increasing the desired cooling power and temperature span increases the amount of magnet that must be used, as expected. It is also seen that the magnetic field is constant as a function of cooling power while it increases monotonically with temperature span. The reason for this behavior is that the only way to increase the temperature span is to increase the magnetic field as $T_\n{span}$ does not depend on $m_\n{mcm}$. Also, note that even for very low temperature spans a magnetic field above 0.6 T is favored. This is because the $M^*$ parameter for a Halbach cylinder drops off significantly at low magnetic fields as shown in Fig. \ref{Fig.Mstar_Halbach}. This is not the case for the $M_{25}$ magnet, where low magnetic fields are favored. From the figure we also see that the amount of magnetocaloric material increases with cooling power, especially for a high value of the temperature span. This can be explained based on Eq. (\ref{Eq.Mstar_Halbach}) which shows that it is too expensive to generate a strong magnetic field and thus it is more favorable to use more magnet material to generate a large cooling power. Finally, note that about two to three times more magnet than magnetocaloric material is used. Therefore the total cost will be roughly proportional to the cost of the magnet.

\subsection{A packed sphere bed regenerator of Gd}
\begin{figure}[!t]
\begin{center}
\subfigure[a]{\includegraphics[width=0.47\textwidth]{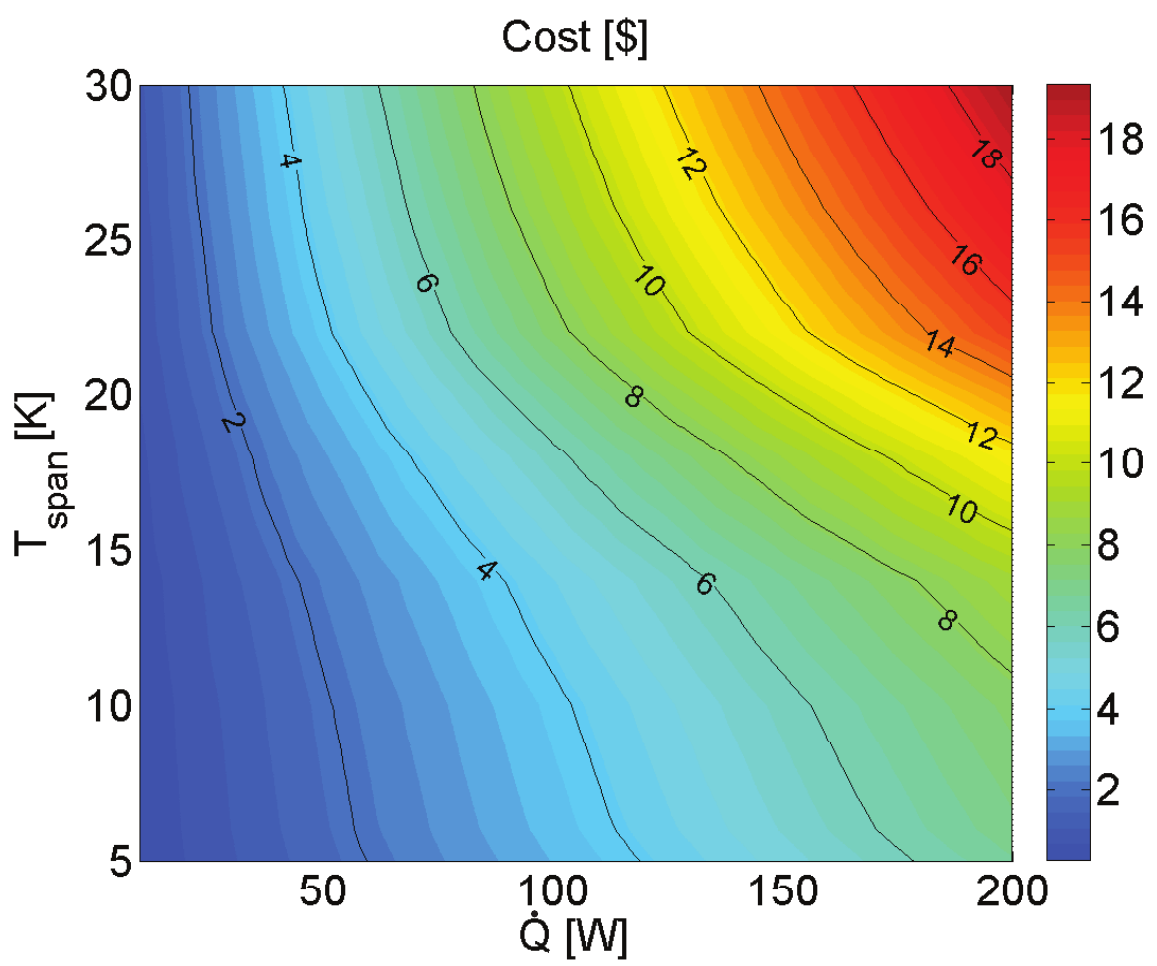}}
\subfigure[b]{\includegraphics[width=0.47\textwidth]{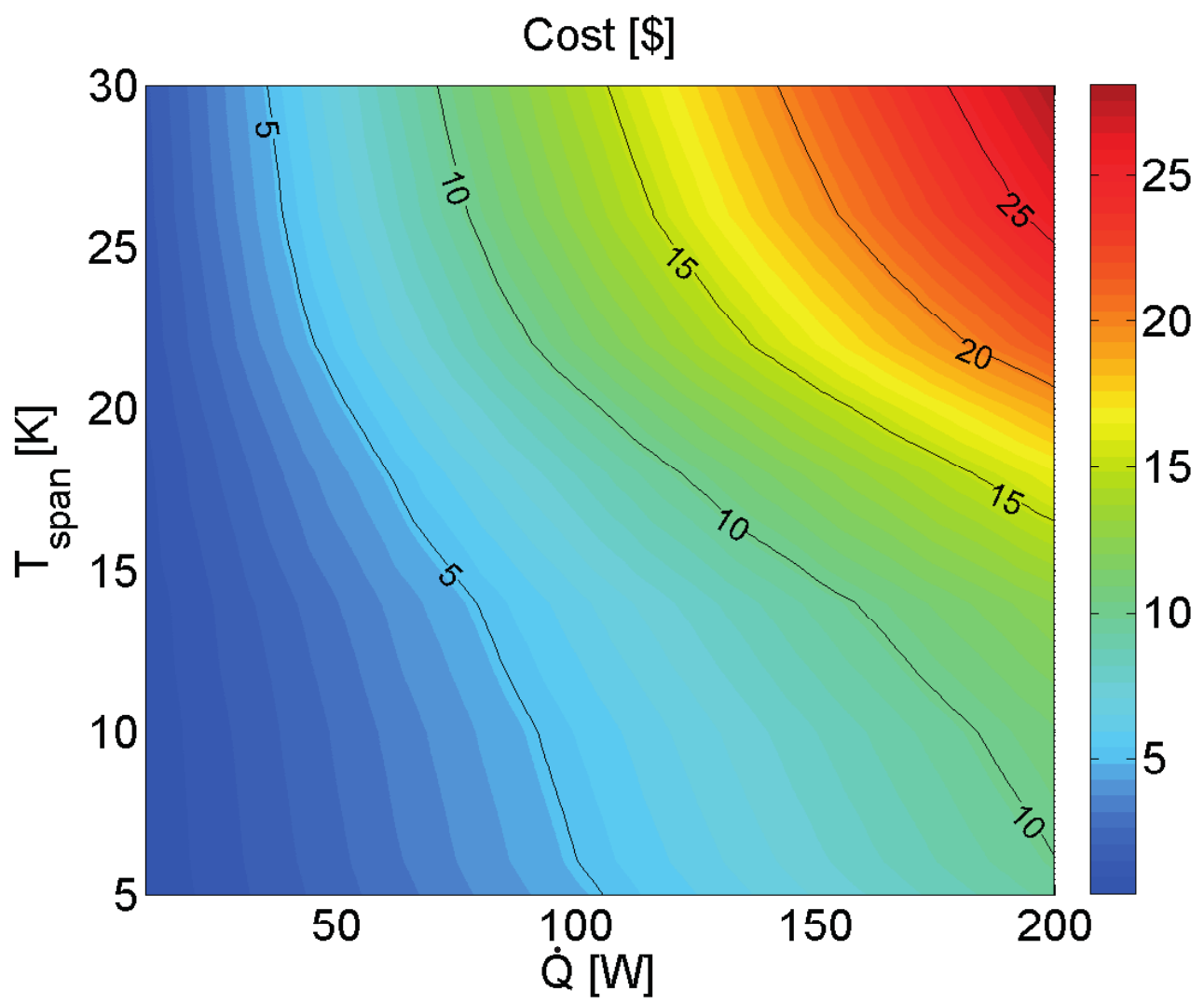}}
\end{center}
\caption{The minimum cost in \$ of a packed sphere bed magnetic refrigeration system of Gd as a function of temperature span and cooling power of an AMR for (a) a $M_{25}$ permanent magnet assembly and (b) a Halbach cylinder of infinite length.}
    \label{Fig.Gd_bed}
\end{figure}

Having considered the cost of the parallel plate regenerator we now consider the packed sphere bed regenerator. In Fig. \ref{Fig.Gd_bed} the minimum combined cost of a packed sphere bed regenerator of Gd and using either an $M_{25}$ or a Halbach magnet is shown. From this figure we see that the total minimum cost of a packed sphere bed regenerator is several times less than that of a parallel plate regenerator. As for the parallel plate case changing from an $M_{25}$ magnet to a Halbach cylinder increases the cost by up to $\sim{}50\%$. Note that we do not consider the COP, i.e. the cost of operating the magnetic refrigerators, but only the cost of the materials for the devices. So although packed sphere beds are better than the parallel plates the energy needed for operation may be higher due to the increased pressure loss.

\begin{figure}[!p]
\begin{center}
\subfigure[a]{\includegraphics[height=0.73\columnwidth]{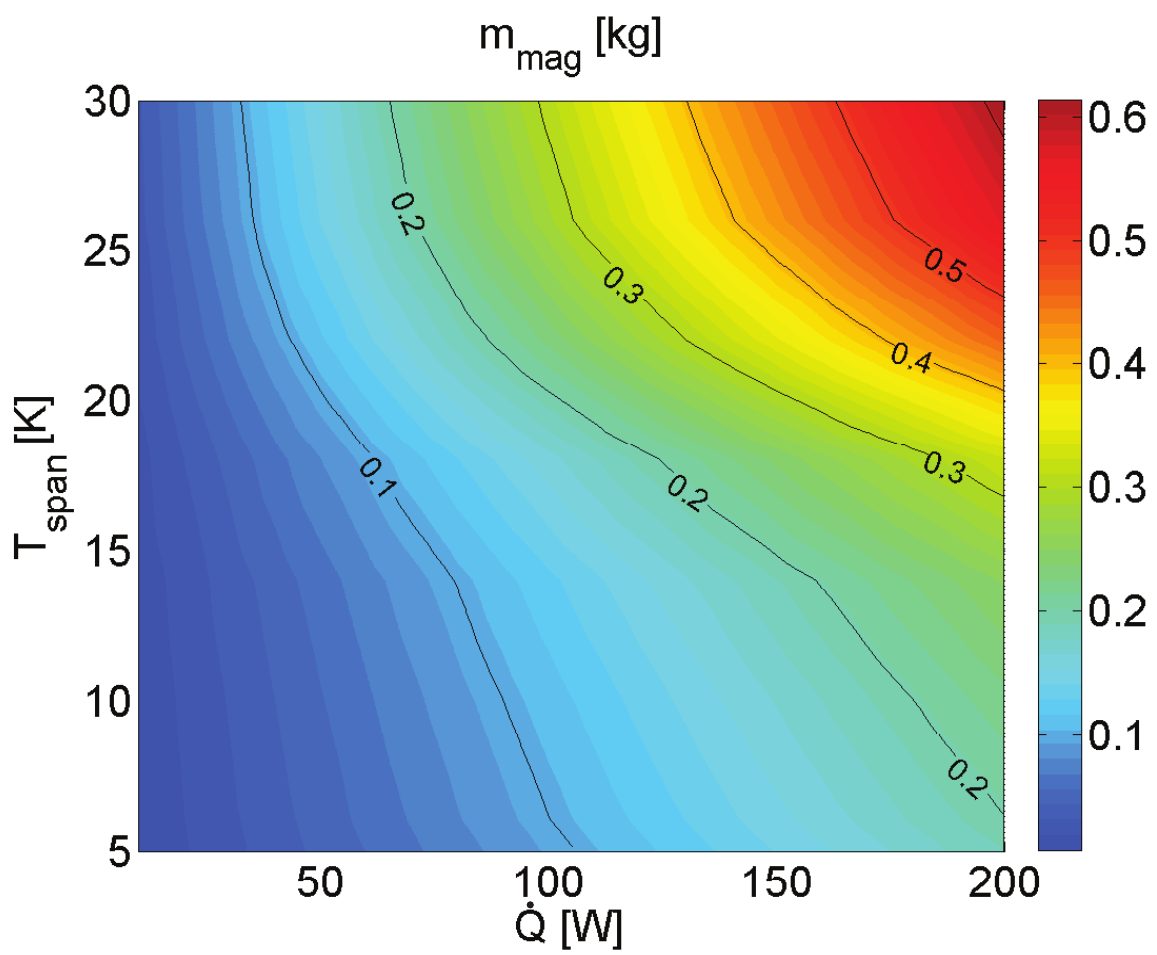}}
\subfigure[b]{\includegraphics[height=0.73\columnwidth]{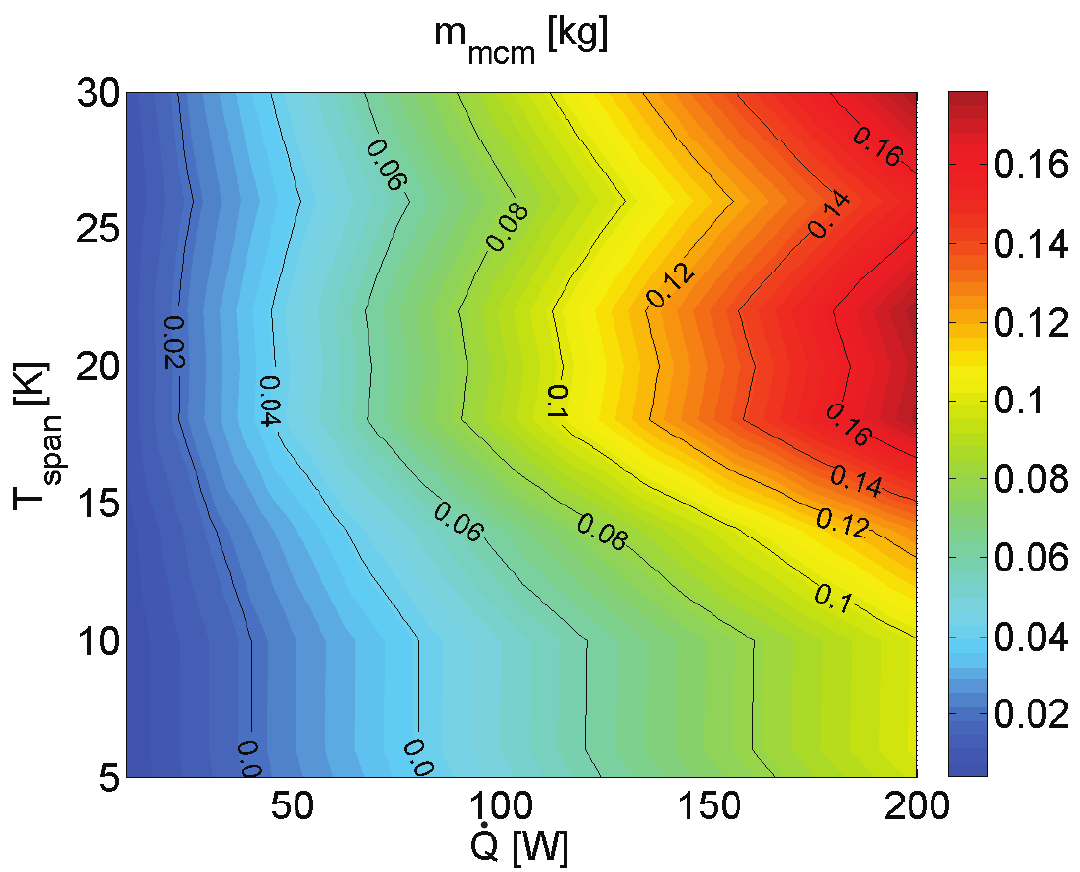}}
\subfigure[c]{\includegraphics[height=0.73\columnwidth]{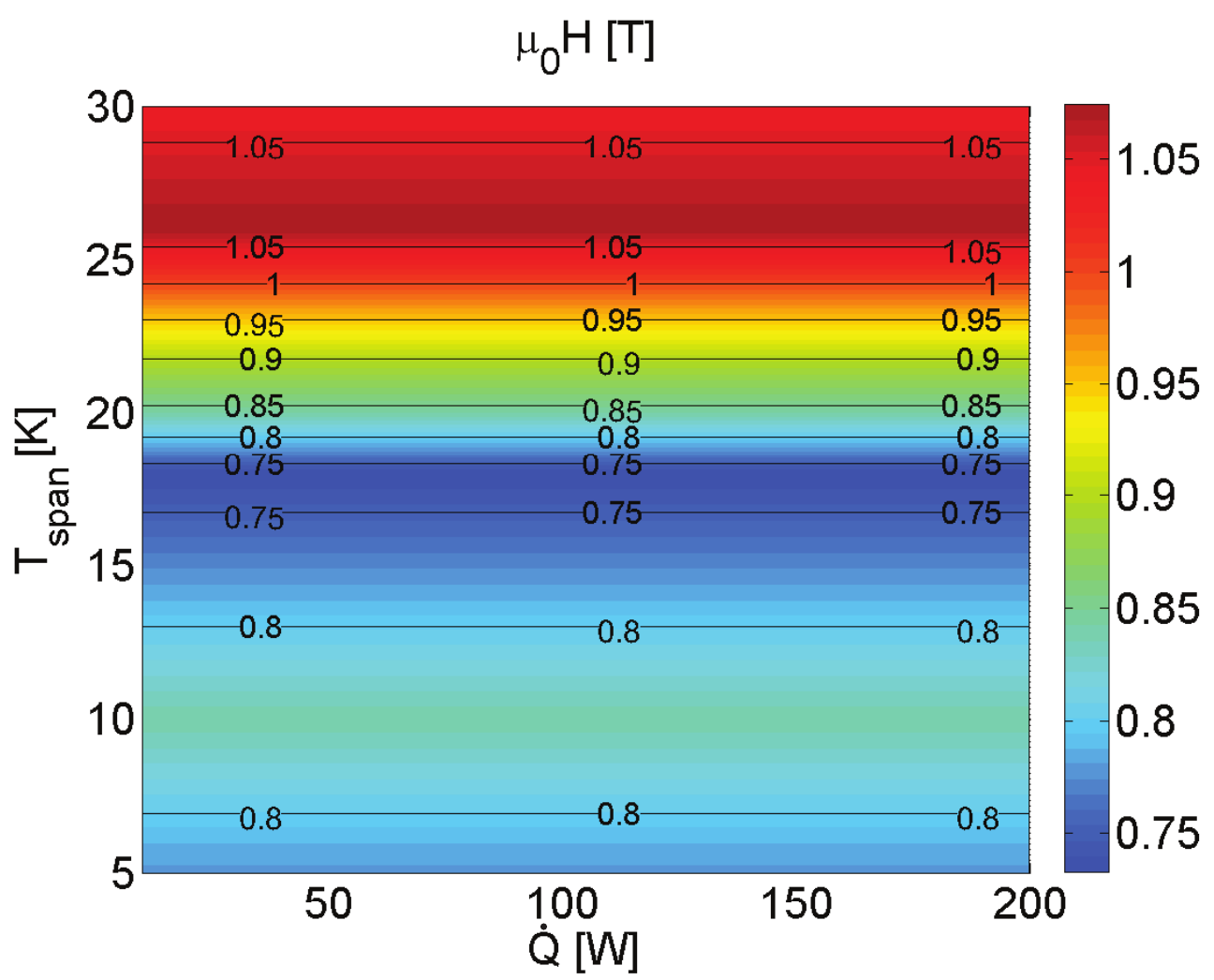}}
\end{center}
\caption{(a) The amount of magnet material, (b) the corresponding amount of magnetocaloric material and (c) the corresponding magnetic field for the magnetic refrigeration system with the lowest combined cost of the magnet and magnetocaloric material for a packed bed AMR with Gd and a Halbach cylinder permanent magnet assembly.}
    \label{Fig.parallel_all_packed}
\end{figure}

In Fig. \ref{Fig.parallel_all_packed} the corresponding amount of magnetocaloric material, magnetic field and magnet material for the minimum combined cost packed sphere bed AMR with Gd and using a Halbach magnet, i.e. the cost shown in Fig. \ref{Fig.Gd_bed}b, is plotted. Compared to the parallel plate case (see Fig. \ref{Fig.Halbach_com}) we see that both the amount of magnet and magnetocaloric material needed has been greatly reduced, which explains the significant reduction in the overall cost of the system. Otherwise, the amount of magnetocaloric material and the value of the magnetic field follow the same trends as seen for the parallel plate case.

\clearpage

\section{A graded magnetocaloric material}
It has previously been shown that a gain in cooling power can be obtained by using several magnetocaloric materials with different Curie temperatures in a so called multimaterial AMR system \citep{Rowe_2006,Jacobs_2009,Nielsen_2010b, Hirano_2010, Russek_2010}. Therefore the cost of a multimaterial AMR will also be considered.

Here we consider a material with a constant $\Delta{}T_\n{ad}$ profile as function of temperature. Such a material can be thought of as representing an infinitely linearly graded AMR for a given temperature span in a AMR operating at steady state. The adiabatic temperature change is chosen to be equal to the peak adiabatic temperature change for commercial grade Gd as reported in \citet{Bjoerk_2010d}, which is $\Delta{}T_\n{ad} = 3.3$ K. The specific heat capacity as a function of temperature in zero applied magnetic field is taken to be constant with a value of $c_\n{p}=270.36$ J kg$^{-1}$ K$^{-1}$, which is the average of the measured value of $c_\n{p}$ for commercial grade Gd in the temperature interval from 220 K to 340 K as reported in \citet{Bjoerk_2010d}. Then the remaining magnetocaloric properties for this material, e.g. the specific heat capacity as a function of magnetic field, are constructed in a thermodynamically consistent way as described in \citet{Engelbrecht_2010b}. Finally the adiabatic temperature change is chosen to scale as a power law with an exponent of 2/3 for all temperatures, i.e. $\Delta{}T_\n{ad}(\mu_{0}H)=\Delta{}T_\n{ad}(1\;\n{T})(\mu_{0}H)^{2/3}$. This is the theoretical scaling of a second order material calculated using mean field theory at the Curie temperature \citep{Oesterreicher_1984} and it has been observed for both Gd and LaFe$_{13-x-y}$Co$_x$Si$_y$ materials \citep{Pecharsky_2006,Bjoerk_2010d}. As the material with the constant adiabatic temperature change is always operating at the Curie temperature the 2/3 power law scaling is assumed to be true for all temperatures. The remaining material properties are taken to be identical to Gd.%, which are the density, $\rho$, and thermal conductivity, $k$, are taken to be constants and have been chosen to be $\rho = 7900$ kg m$^{-3}$ and $k = 10.5$ W m$^{-1}$ K$^{-1}$, which are the values for Gd.

\begin{figure*}[!t]
\begin{center}
\subfigure[a]{\includegraphics[width=0.47\textwidth]{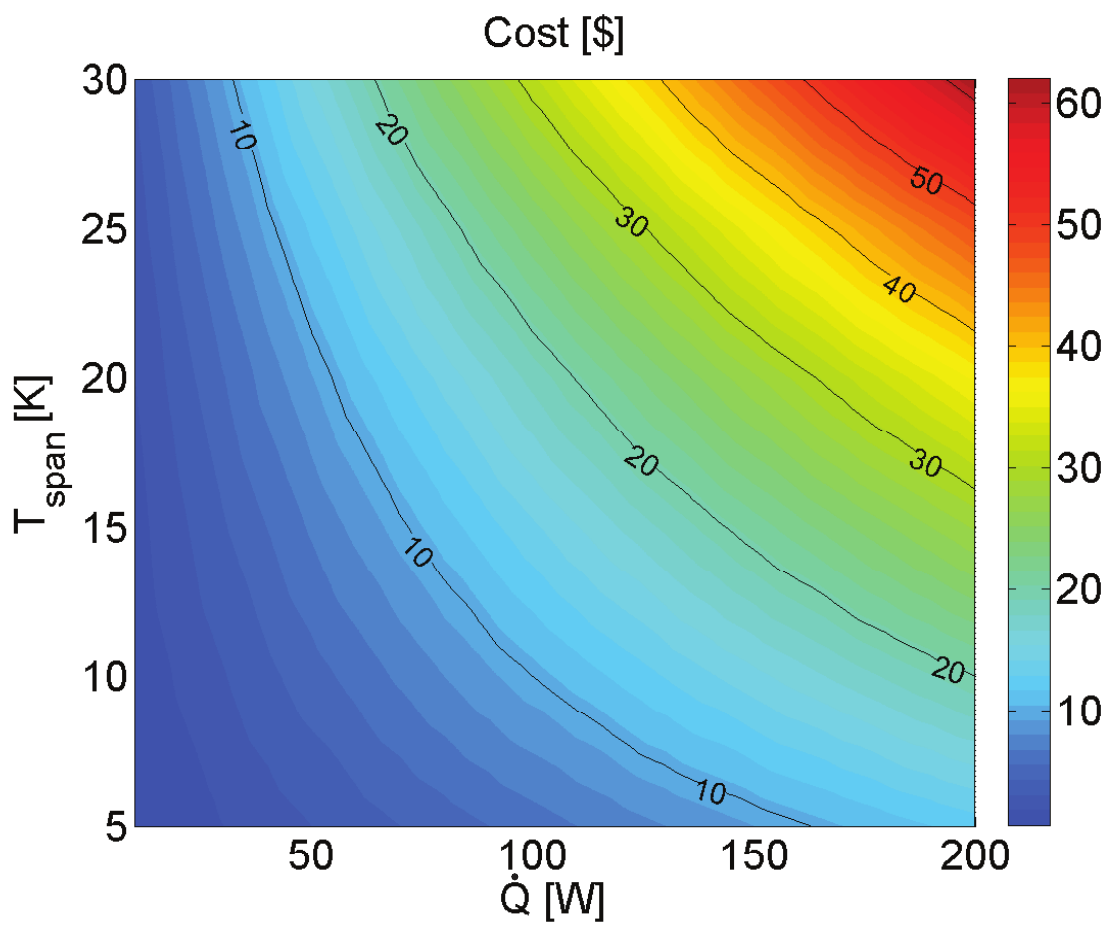}}
\subfigure[b]{\includegraphics[width=0.47\textwidth]{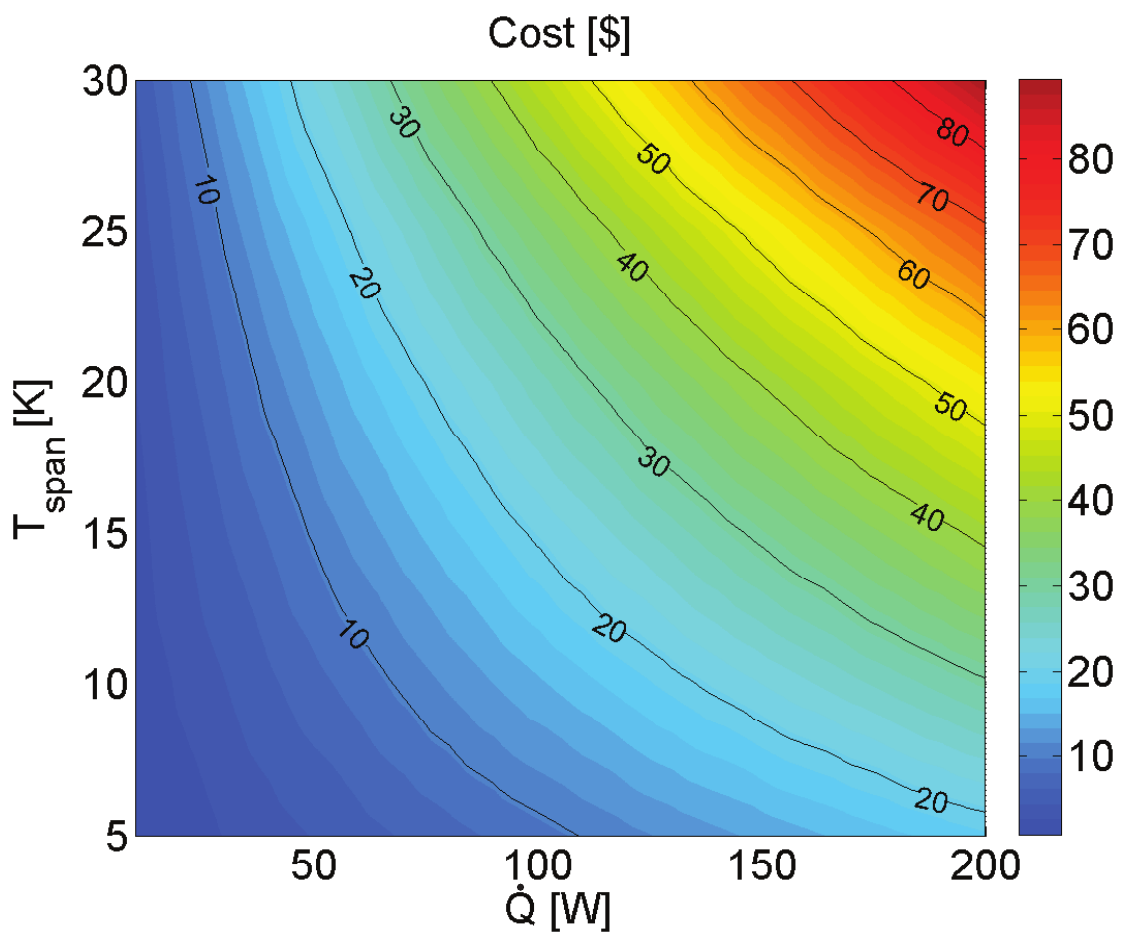}}
\end{center}
\caption{The minimum combined cost in \$ of a parallel plate magnetic refrigeration system with a magnetocaloric material with constant $\Delta{}T_\n{ad}$ as a function of temperature span and cooling power for (a) a $M_{25}$ permanent magnet assembly and (b) a Halbach cylinder of infinite length.}
    \label{Fig.Flat_const}
\end{figure*}

\begin{figure*}[!t]
\begin{center}
\subfigure[a]{\includegraphics[width=0.47\textwidth]{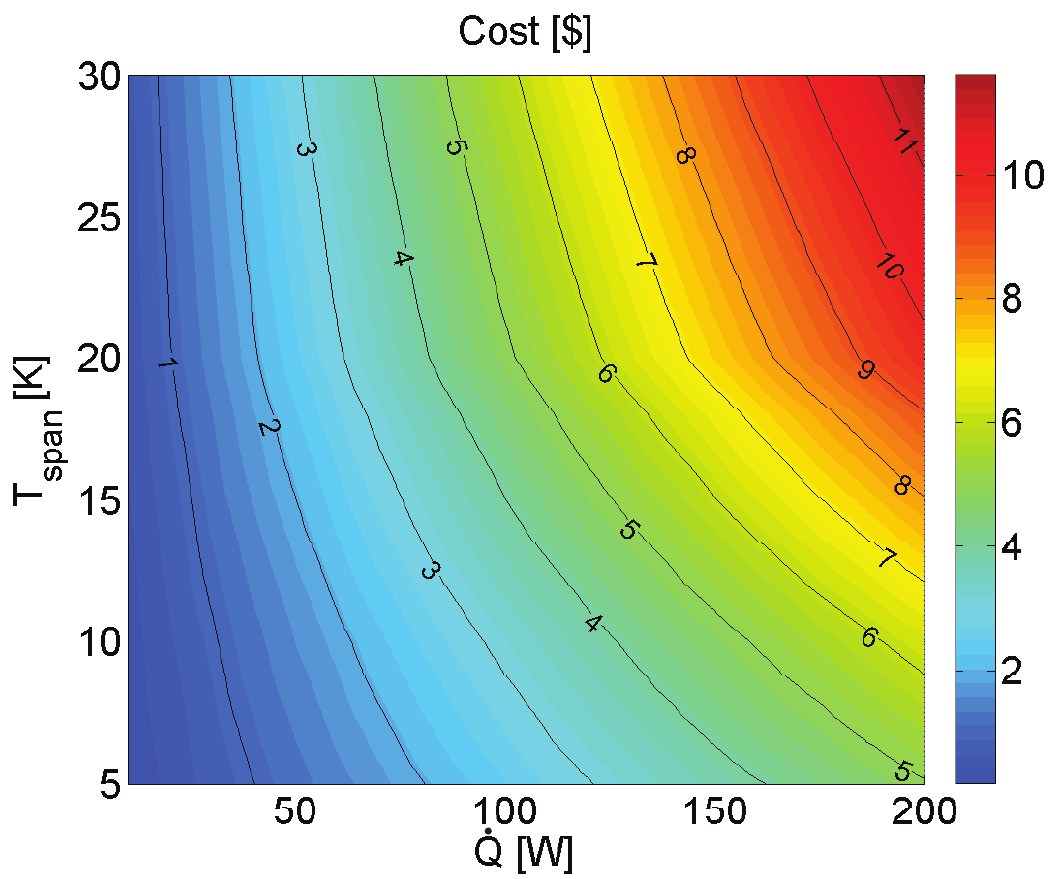}}
\subfigure[b]{\includegraphics[width=0.47\textwidth]{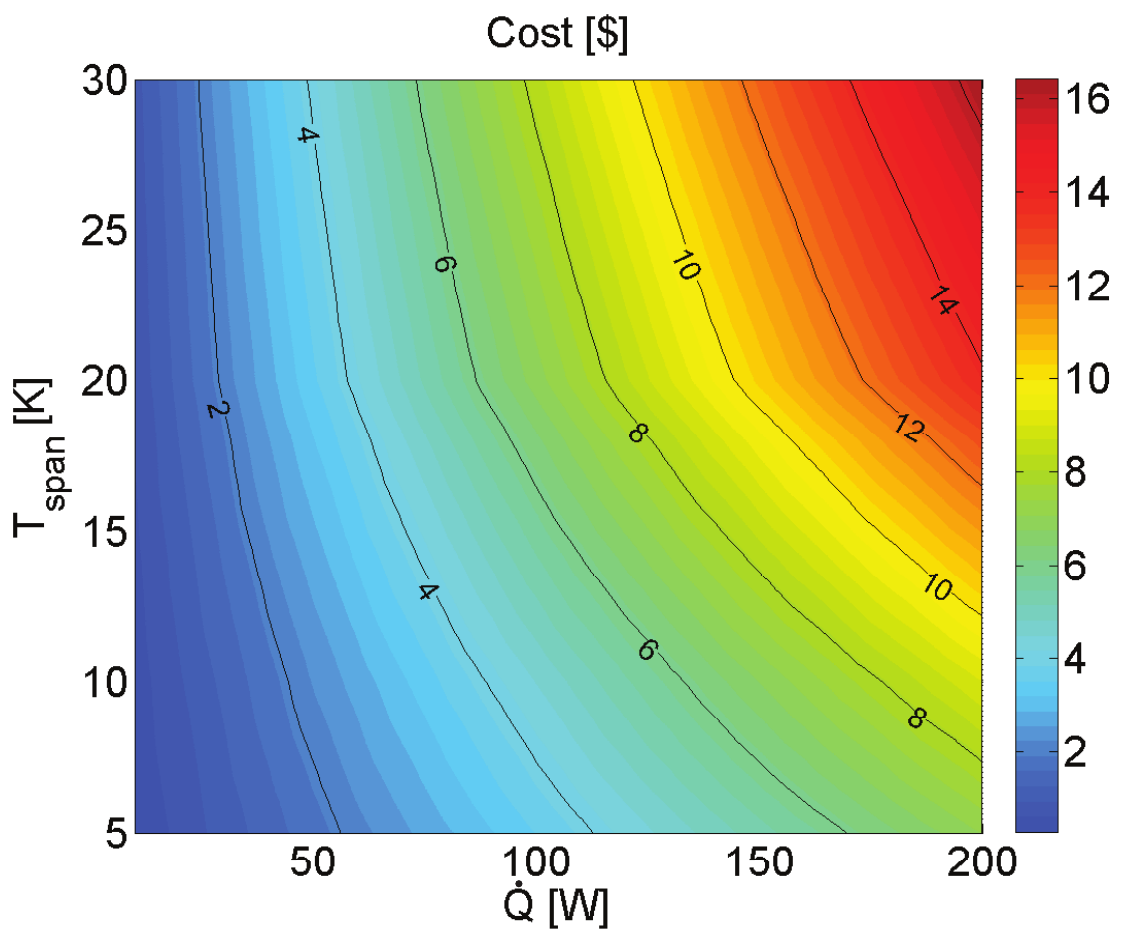}}
\end{center}
\caption{The minimum combined cost in \$ of a packed sphere bed magnetic refrigeration system with a constant $\Delta{}T_\n{ad}$ magnetocaloric material as a function of temperature span and cooling power for (a) a $M_{25}$ permanent magnet assembly and (b) a Halbach cylinder of infinite length.}
    \label{Fig.Flat_bed_const}
\end{figure*}

The cooling power has been computed for all process parameters given in Table \ref{Table.Packed_bed} and \ref{Table.Parallel_plate}, as for Gd. Shown in Fig. \ref{Fig.Flat_const} is the minimum combined cost for a magnetic refrigeration system using the constant $\Delta{}T_\n{ad}$ material, for both an $M_{25}$ and a Halbach magnet. Compared to the device using Gd (see Fig. \ref{Fig.cost}) it is seen that using a material with constant $\Delta{}T_\n{ad}$ can optimally reduce the cost of the refrigeration device by 50\%. The same conclusions apply to Fig. \ref{Fig.Flat_bed_const} which shows the minimum combined cost for the packed bed system using a constant $\Delta{}T_\n{ad}$ material. %Again, note that the operating cost of a packed bed AMR is most likely significantly higher than for a parallel plate AMR as the pressure drop is likely orders of magnitude higher for the packed bed case.

\begin{figure}[!p]
\begin{center}
\subfigure[a]{\includegraphics[height=0.73\columnwidth]{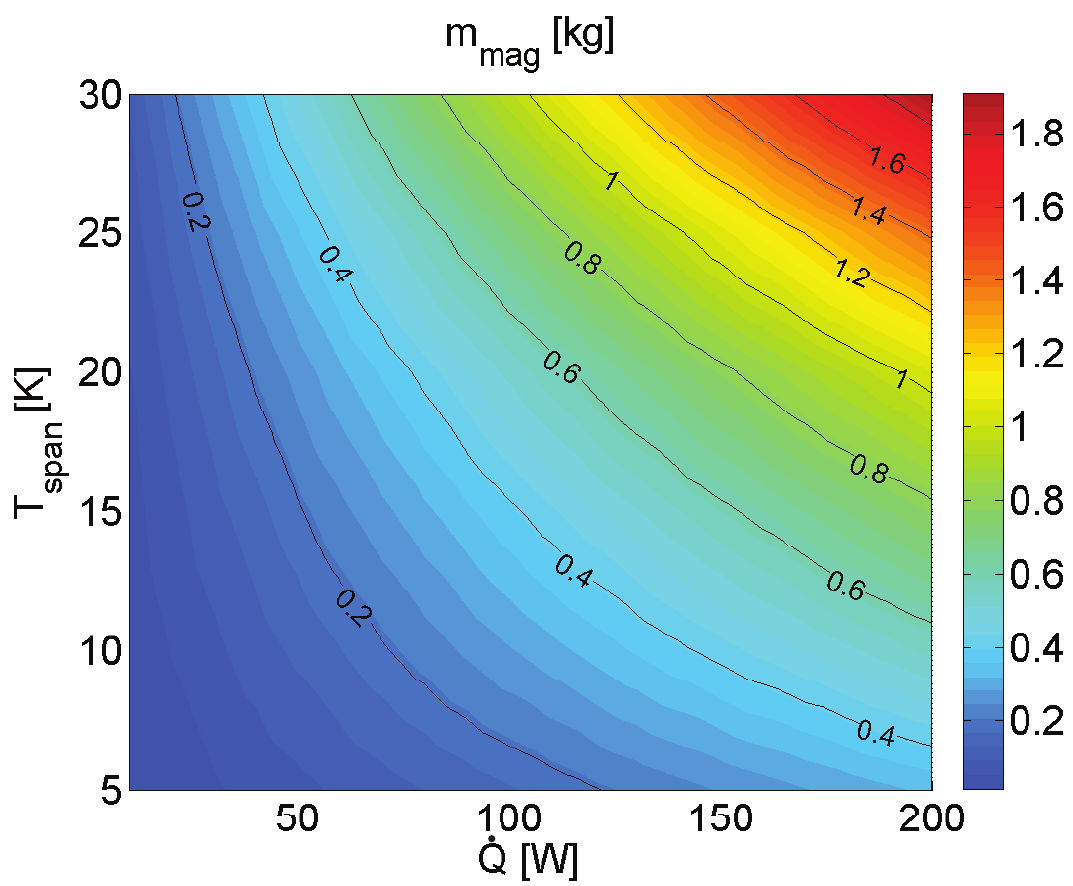}}
\subfigure[b]{\includegraphics[height=0.73\columnwidth]{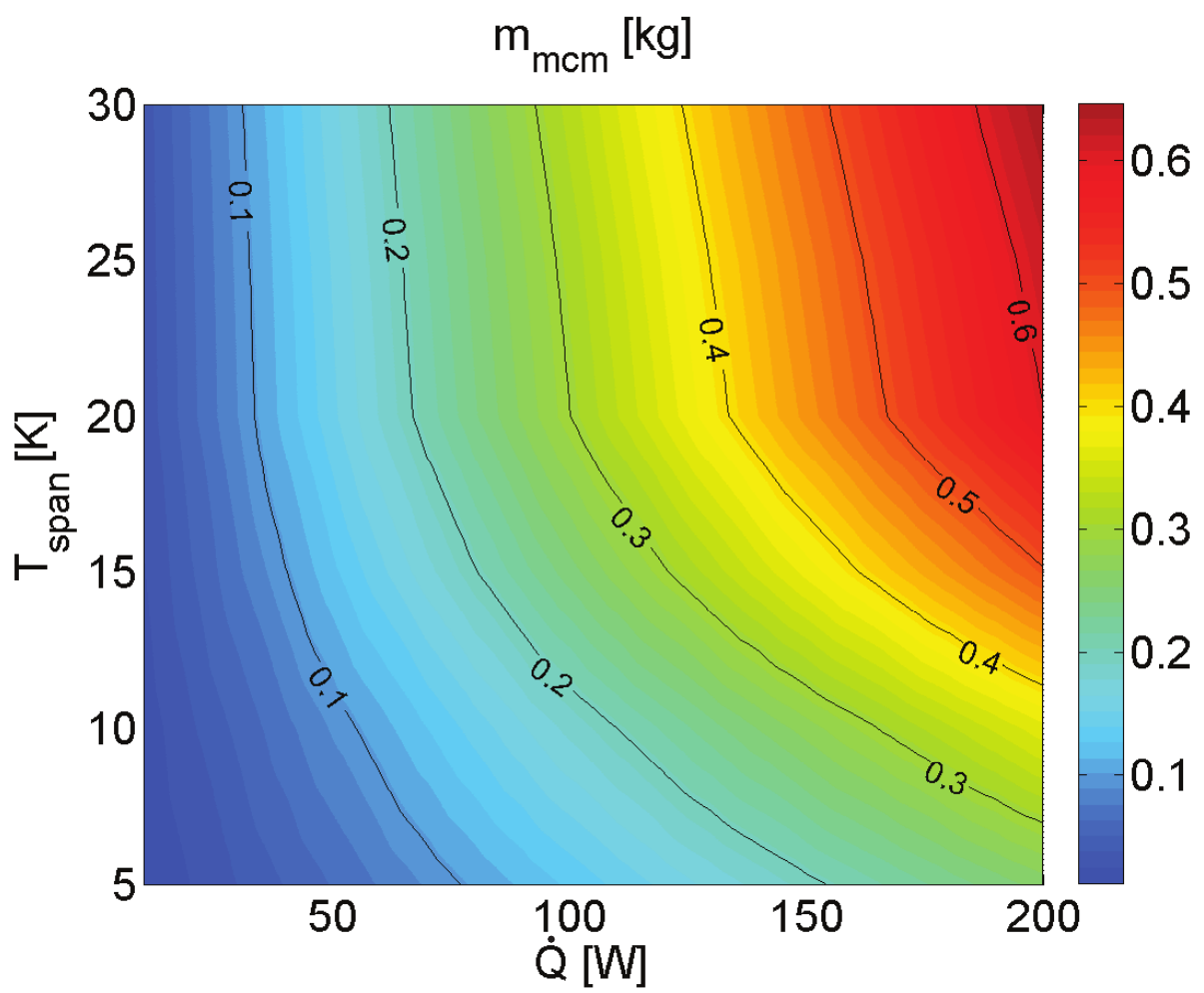}}
\subfigure[c]{\includegraphics[height=0.73\columnwidth]{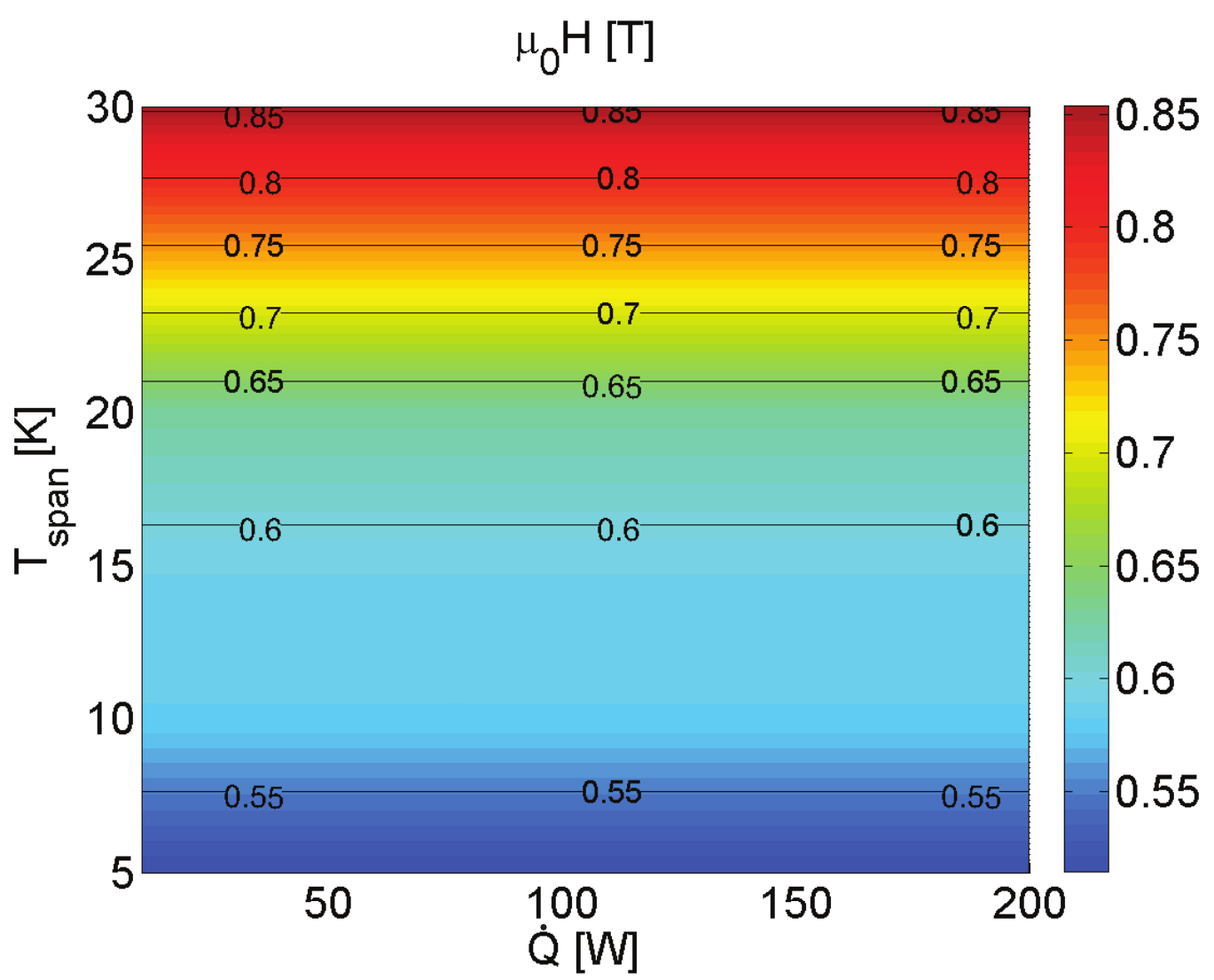}}
\end{center}
\caption{(a) The amount of magnet material, (b) the corresponding amount of magnetocaloric material and (c) the corresponding magnetic field for the magnetic refrigeration system with the lowest combined cost of the magnet and magnetocaloric material for a parallel plate constant $\Delta{}T_\n{ad}$ AMR with a Halbach cylinder permanent magnet assembly.}
    \label{Fig.parallel_all_const}
\end{figure}

Shown in Fig. \ref{Fig.parallel_all_const} is the corresponding amount of magnetocaloric material, magnetic field and magnet material for the minimum combined cost parallel plate AMR with a constant $\Delta{}T_\n{ad}$ material and using a Halbach magnet, cf. Fig. \ref{Fig.Flat_const}b. Compared to Fig. \ref{Fig.Halbach_com} the amount of magnet material needed has been halved, which is also the reason for the overall reduction in cost. Even though the cooling power of the constant $\Delta{}T_\n{ad}$ material for most temperatures is higher than for Gd it is seen that the same amount of magnetocaloric material is used in the AMR. However, the value of the magnetic field has been reduced by up to $\sim{}0.3$ T compared to Fig. \ref{Fig.Halbach_com}. The reason for this is that much more magnet material is used than magnetocaloric material and as the magnet material is twice as expensive as the magnetocaloric material it is much more advantageous to reduce the amount of magnet material that is used, thereby lowering the value of the magnetic field.

\section{Discussion}
The cost determined above is only the cost of the permanent magnet material and the magnetocaloric material that go into the magnetic refrigeration system. Additional costs such as motors, pump etc. needed for the refrigeration system have not been included in the analysis. Especially for the packed sphere bed regenerator the pressure drop across the AMR may be high and this will introduce a significant energy consumption and thus operating cost. This aspect has not been included in the analysis presented here, but could easily be included if the ``price per pressure drop'' is known or the true operating cost could be estimated. Also, as previously mentioned, the process parameters that produce the lowest combined cost do not necessarily represent the process parameters with highest COP. This implies that the device with highest COP will be more expensive to produce then the minimum cost found here. Finally, the cooling power is computed using a numerical model which have been shown to produce higher cooling capacities than seen experimentally \citep{Engelbrecht_2008}, due to experimental heat losses. Also, demagnetization effects have not been taken into account, which will lower the internal magnetic field in the magnetocaloric material. Therefore a larger magnetic field or more magnetocaloric material might be needed than found above, which will increase the total cost of the system.

The material with constant adiabatic temperature change, used here to represent an infinitely linearly graded AMR, might also not represent the optimal grading of an AMR. The ideal grading will depend on the magnetocaloric materials used, and thus a more efficient grading and thereby a lower cost device might be possible. Also, using magnetocaloric materials with a higher adiabatic temperature change will lower the cost of the device, as more cooling power can be produced per mass of magnetocaloric material. Note that the calculations of the cost assume that the magnet is utilized, i.e. filled with magnetocaloric material, at all times. If this is not the case, the system will become more expensive. Nevertheless, although this analysis as mentioned above contain several assumptions, the estimation of the cost based on the prediction of the model is still valid for a large number of process parameters and as a reasonable minimum cost of a magnetic refrigeration system.

\begin{figure}[!t]
\begin{center}
\includegraphics[width=1\columnwidth]{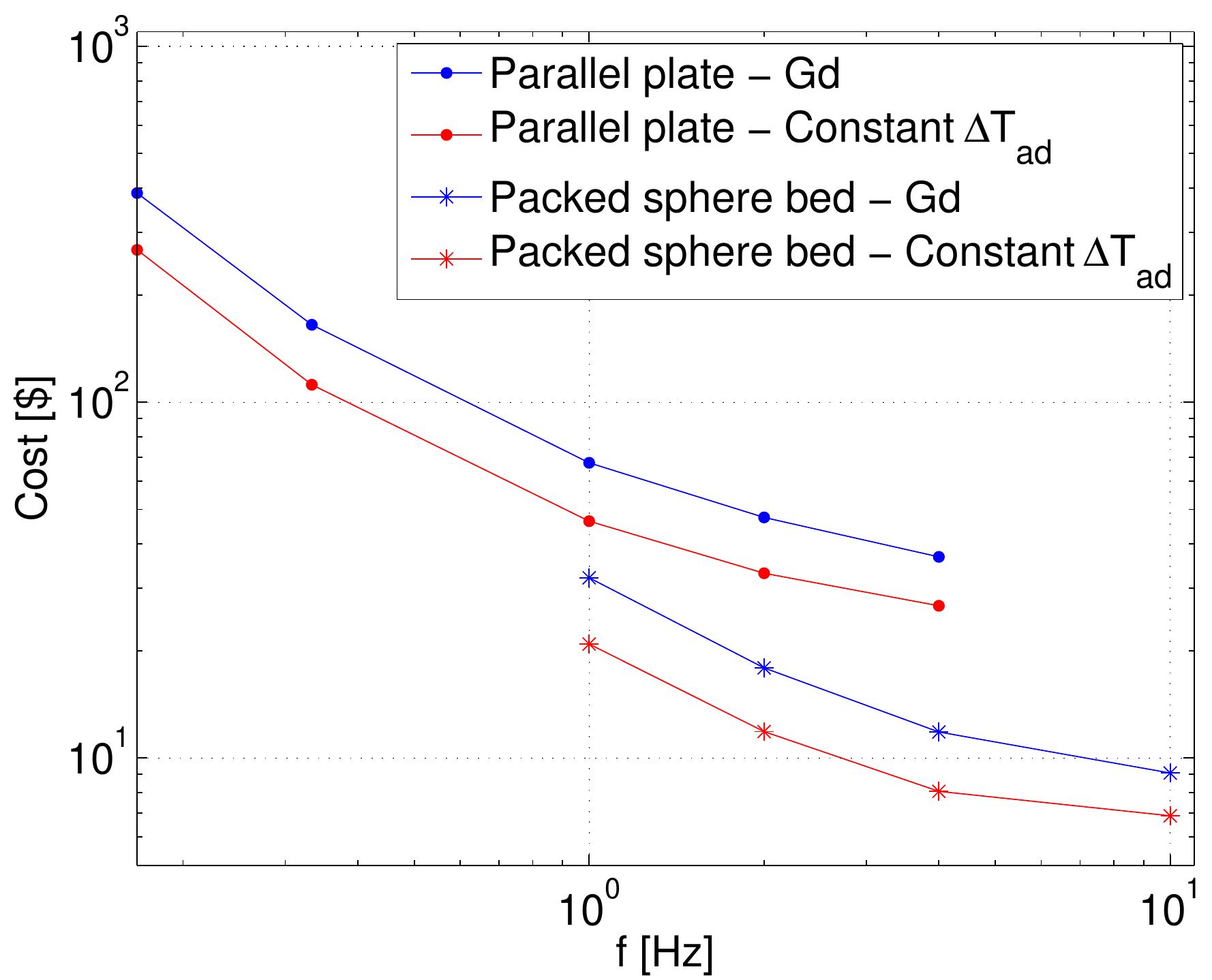}
\end{center}
\caption{The minimum combined cost as a function of frequency for a magnetic refrigeration device using a Halbach magnet and with a temperature span of 20 K and a cooling power of 100 W. The lines are only meant to serve as guides to the eye.}
    \label{Fig.Frequency_cost}
\end{figure}

It is of interest to consider the reduction in cost that is possible by increasing the operating frequency, $f$, of the system. Shown in Fig. \ref{Fig.Frequency_cost} is the minimum combined cost of a magnetic refrigerator with a Halbach magnet and with a temperature span of 20 K and a cooling power of 100 W as a function of frequency. These numbers represent reasonable operating parameters of a magnetic refrigerator.

It can clearly be seen that increasing the frequency significantly reduces the cost of the refrigerator. No optimum in frequency is seen, but this might be caused by the limited choice of process parameters studied. However, note that the operating cost will also depend on the frequency. Besides increasing the frequency the cost found in the above analysis can also be lowered by examining other process parameters than those considered here or by considering alternative regenerator geometries where the thermal contact between the heat transfer fluid and the regenerator is higher but the pressure drop is not excessive.

\begin{figure}[!t]
\begin{center}
\subfigure[a]{\includegraphics[height=0.73\columnwidth]{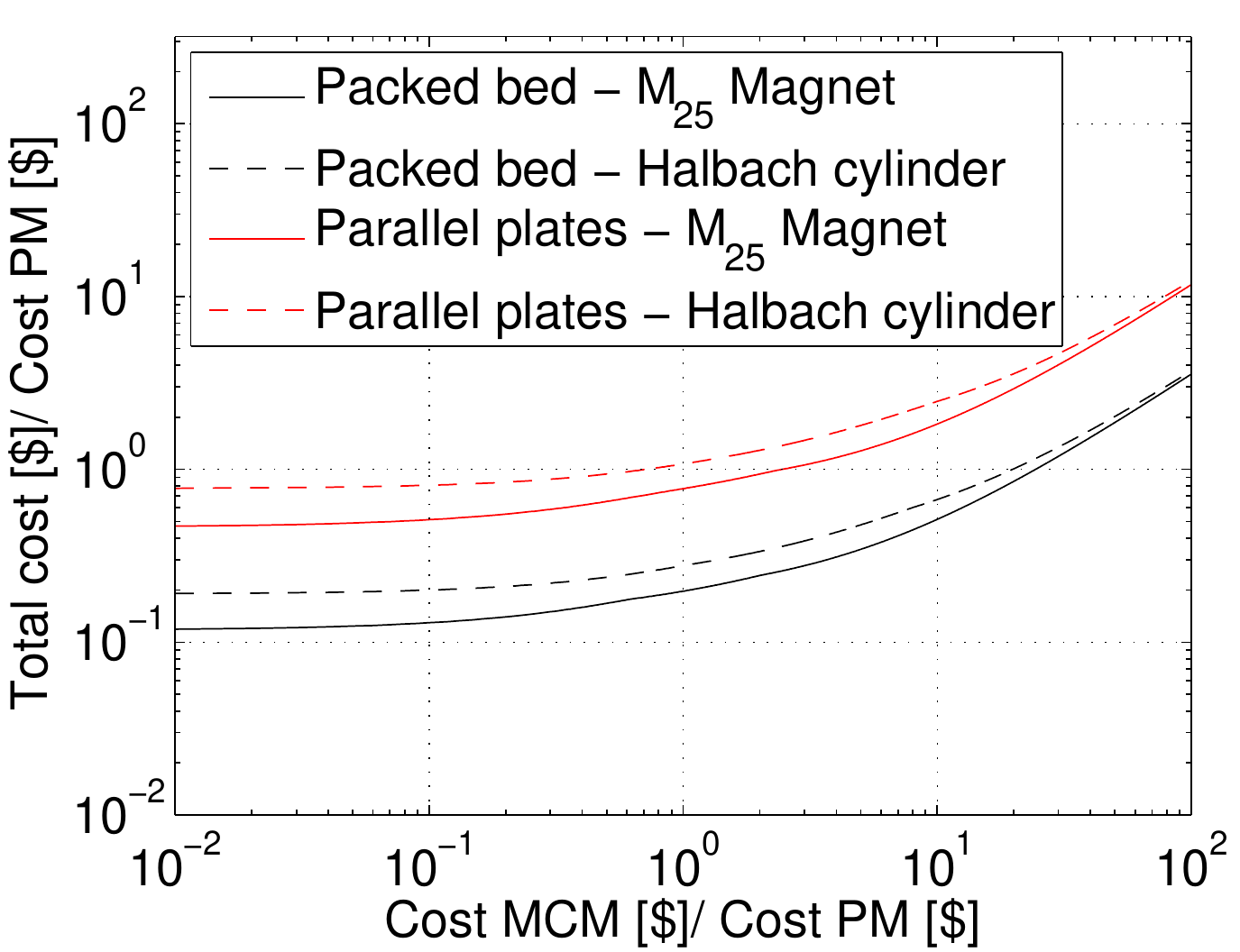}}
\subfigure[b]{\includegraphics[height=0.73\columnwidth]{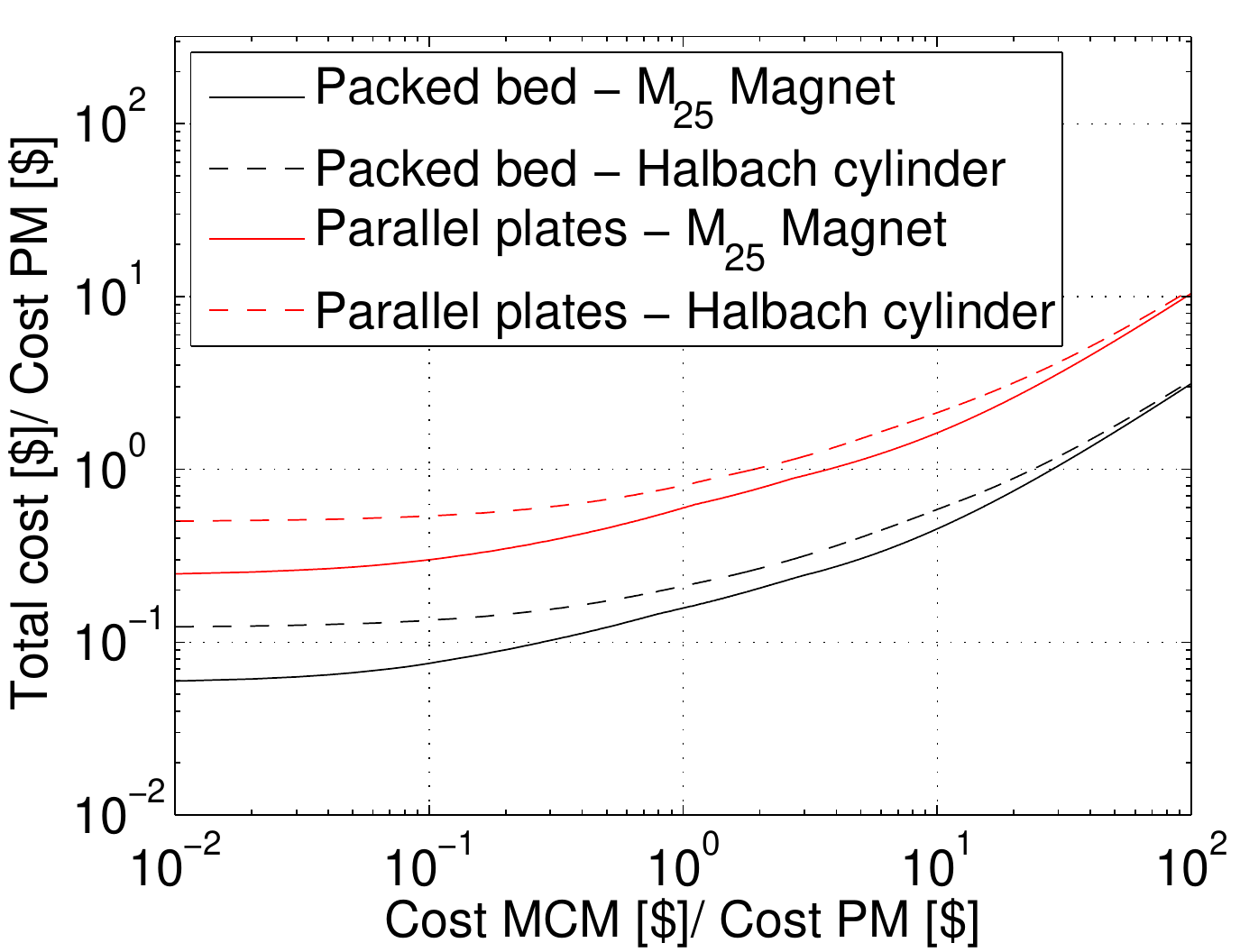}}
\end{center}
\caption{The ratio of the minimum total cost and the cost of the permanent magnet (PM) material as a function of the ratio between the cost of the magnetocaloric material (MCM), and the permanent magnet material for different AMRs but all with a temperature span of 20 K and a cooling power of 100 W. The magnetocaloric material in (a) is taken to be Gd while in (b) it is the constant $\Delta{}T_\n{ad}$ material.}
    \label{Fig.Cost_relative}
\end{figure}

In the present study the cost of permanent magnet (PM) material was assumed to be \$40 per kg and for the magnetocaloric material (MCM) the cost was \$20 per kg. However, it is of interest to investigate the total cost of the AMR for different costs of the magnet and magnetocaloric material. Shown in Fig. \ref{Fig.Cost_relative} is the minimum total cost of several different types of AMR with a temperature span of 20 K and a cooling power of 100 W as a function of the ratio of the cost of the magnet and magnetocaloric material. As can be seen from the figure the minimum total cost for all AMRs behave in much the same way. For a small ratio of Cost MCM / Cost PM the total cost scales linearly with the cost of the magnet as the price of the magnetocaloric material is small compared to the price of magnet and thus can be ignored. It is also seen that it is always better to use an $M_{25}$ magnet compared to a Halbach magnet, as expected. Using this figure the minimum cost of a given AMR can be calculated for any cost of the magnet and magnetocaloric materials.

\section{Conclusion}
In this paper an expression is proposed for the total mass and thus cost of the magnet material and magnetocaloric material needed to construct a magnetic refrigerator. It is shown that for equal densities of the magnet and magnetocaloric material and a magnetic field equal to the remanence and a system with a porosity of 0.5 that the magnet with the largest possible figure of merit, termed the $M_{25}$ magnet, must have a mass at least four times the mass of magnetocaloric material used, if the magnet is used at all times. For a Halbach cylinder the mass of the magnet is even larger.

The total minimum mass and cost of both a parallel plate and packed sphere bed regenerator consisting of the magnetocaloric material Gd or a material with a constant adiabatic temperature change profile was also studied. Using the cooling power computed from a numerical model for 15876 packed sphere bed process parameters and for 14994 parallel plate process parameters, all realistically chosen, the minimum cost of such regenerators was estimated. This was done for both an $M_{25}$ magnet and for a Halbach cylinder of infinite length. The cost, amount of magnet and magnetocaloric material as well as the magnetic field was determined as functions of desired temperature span and cooling power for the cheapest overall design. Assuming a cost of magnet material of \$40 per kg and of magnetocaloric material of \$20 per kg the cheapest parallel plate refrigerator with Gd that produces 100 W of continuous cooling at a temperature span of 20 K using a Halbach magnet will use around 0.8 kg of magnet, 0.3 kg of Gd, have a magnetic field of 0.8 T and have a minimum cost of \$35. The cost is dominated by the cost of the magnet. Using a magnetocaloric material with a constant adiabatic temperature profile reduces this cost to \$25 while using a packed sphere bed, also of a constant magnetocaloric material, brings the cost down to \$7. It was also shown that the cost can be reduced by increasing the frequency of the AMR. Finally, the lowest cost was also found as a general function of the cost of the magnet and magnetocaloric material.

\section*{Acknowledgments}
The authors would like to acknowledge the support of the Programme Commission on Energy and Environment (EnMi) (Contract No. 2104-06-0032) which is part of the Danish Council for Strategic Research.

\bibliographystyle{elsarticle-harv}

\end{document}